\documentclass{aa}
\usepackage{endnotes}

\usepackage{booktabs}
\usepackage{graphicx}
\usepackage{comment}
\usepackage{appendix}
\usepackage{placeins}
\usepackage{enumitem}
\usepackage[breaklinks=true]{hyperref}
\usepackage{url}
\usepackage{txfonts}
\usepackage{xcolor}

\begin{document} 

   \title{Catalogue and statistics of >100 MeV solar proton events during solar cycles 23–25 from SOHO/ERNE observations}

   \titlerunning{Catalogue and statistics of >100 MeV solar proton events}

   \subtitle{}

   \author{M. Jarry\inst{1}
          \and
          C. Palmroos\inst{2}
          \and
          E. Lavasa\inst{1}
          \and
          N. Talebpour Sheshvan\inst{3}
          \and
          M. Koeberle\inst{4}
          \and
          B. Heber\inst{4}
          \and
          A. P. Rouillard\inst{3,5}
          \and
          A. Papaioannou\inst{1}
          \and
          J. Gieseler\inst{2}
          \and
          C. Ngom\inst{2}
          \and
          P. Oleynik\inst{2}
          \and
          E. Riihonen\inst{2}
          \and
          R. Vainio\inst{2}
          \and
          G. Vasalos\inst{1}
          \and
          A. Anastasiadis\inst{1}
          }

   \institute{
         Institute for Astronomy, Astrophysics, Space Applications and Remote Sensing (IAASARS), National Observatory of Athens, I. Metaxa \& Vas. Pavlou St., 15236 Penteli, Greece\\ \email{mjarry@noa.gr}
         \and
         Department of Physics and Astronomy, University of Turku, Turku, Finland
         \and
         Institut de Recherche en Astrophysique et Plan{\'e}tologie (IRAP), CNRS, Universit{\'e} de Toulouse, France
         \and
          Institut für Experimentelle und Angewandte Physik, Christian-Albrechts-Universität zu Kiel, 24118 Kiel, Germany 
          \and
          Leibniz-Institut für Astrophysik Potsdam, Potsdam, 14482, Germany
         }

   \date{Received 4 November 2025 / Accepted 24 February 2026}
 
  \abstract
   {
    The SPEARHEAD (specification, analysis, and re-calibration of high-energy particle data) project, funded by the European Union’s Horizon Europe programme, enhances the accuracy and usability of high-energy particle measurements.
    It investigates particle acceleration, release, and transport during solar eruptions by refining instrument response functions and cross-calibrating datasets.
    This effort supports in-depth analyses of solar energetic particle (SEP) events and provides public data products.
   }
   {
   We present a comprehensive catalogue of $>$100~MeV proton events identified from May~1996 to August~2024 using SOHO/ERNE penetrating-particle rates, together with associated solar phenomena derived from multi-instrument observations.
   }
   {
    The SEP events were detected through a systematic scan of ERNE/HED counter data and cross-calibrated with SOHO/EPHIN to derive peak fluxes and fluences.
    Each event was associated with its likely parent eruption using X-ray (XR) (GOES/XRS, RHESSI, SolO/STIX), radio (Wind/WAVES, STEREO/WAVES, ground-based observatories), and $\gamma$-ray (Fermi/LAT) observations, CMEs (SOHO/LASCO), and ground-level enhancements (GLEs) (neutron monitors).
    Timings and physical properties were systematically compared to investigate the relationships between SEP onset, flare evolution, CME kinematics, and radio signatures.
   }
   {
    Associations were established for 96\% of CMEs and 76\% of X-ray flares; all events exhibited type~III radio bursts, while 95\% had type~II, and 53\% type~IV.
    $>$100~MeV $\gamma$-rays were identified in 65 events, and 22 were linked to GLEs or sub-GLEs.
    Statistical analyses reveal that most SEP releases closely follow flare and CME onsets, with moderate SEP-XR-CME correlations, and a strong SEP-GLE fluence link.
    These results indicate that high-energy SEP events are typically associated with strong solar activity signatures, with the observed intensities and timings strongly modulated by magnetic connectivity and coronal conditions.
   }
   {
   This catalogue provides the most extensive reference to date for high-energy SEP events over solar cycles~23–25, establishing a unified framework for future investigations of extreme particle acceleration.
   }

   \keywords{Solar energetic particles (SEPs) -- coronal mass ejections (CMEs) -- solar flares
               }

   \maketitle

\section{Introduction}

Solar energetic particles (SEPs) are high-energy ions and electrons accelerated during solar eruptive events, such as flares and coronal mass ejections (CMEs).
Among them, high-energy SEP events implying protons with energies above 100 MeV are of particular interest due to their ability to reach Earth's atmosphere, posing potential hazards to space-borne and ground-based technologies.
They represent the upper end of the SEP energy distribution, and tend to be associated with faster CMEs, more intense soft X-ray flares, and earlier particle release, often during the flare rise phase \citep[e.g.][]{Kahler_1984, Cliver_2004, Cane_2010, Gopalswamy_2012, Reames_2013, Richardson_2014, Dierckxsens_2015, Trottet_2015}.
Recent analysis by \citet{Ameri_2024} showed that these features distinguish them from lower-energy SEP events and point to low-coronal acceleration mechanisms.
Recent work by \citet{Ameri_2024} provided a systematic, energy-resolved confirmation and quantification of these established trends, showing that these characteristics correlate with SEP event energy.
At the same time, acceleration efficiency, and thus the occurrence of high-energy SEPs also depends on shock properties, coronal magnetic field configuration, and pre-existing turbulence or seed populations.
Despite decades of observations, several aspects of their acceleration and transport remain uncertain, particularly the relative contributions of processes associated with flares and CME-driven shocks.
The Energetic and Relativistic Nuclei and Electron experiment \citep[ERNE;][]{Torsti_1995} on board the Solar and Heliospheric Observatory \citep[SOHO;][]{Domingo_1995} spacecraft has been monitoring solar particles near Earth since 1996, providing a unique long-term dataset of high-energy proton measurements.
While the scientific data products of ERNE's High Energy Detector (HED) provide reliable measurements of proton intensities up to energies of 50 MeV/n, a previously unused on-board counter detecting particles penetrating the instrument has been investigated for its scientific research potential within the SPEARHEAD project. This analysis has not been finished yet, so these observations are so far only available as count rates, but the effective energy has already been determined to be about 130 MeV.
Several other instruments have provided high-energy proton measurements over the past decades, including IMP-8/GME \citep[e.g.][]{Richardson_2008}, SAMPEX \citep[e.g.][]{Mewaldt_2012}, and the PAMELA mission, which reported detailed spectra of numerous $>$80 MeV SEP events during solar cycle 24 \citep[e.g.][]{Bruno_2018}.
These complementary datasets place the ERNE observations in a broader context of high-energy SEP measurements.

In this paper, we present a catalogue of $>$100 MeV proton events based on the scanning of SOHO/ERNE/HED measurements (for the first time using particles penetrating the instrument at an energy above 100 MeV), complemented with SEP characteristics from SOHO/Electron Proton Helium Instrument \citep[EPHIN;][]{Mueller-Mellin_1995} and with comprehensive solar associations.
The time coverage extends from May 1996 to 2024 (for the first version of the catalogue), meaning complete coverage of solar cycles 23 and 24, and it also covers the ongoing solar cycle 25.
A total of 172 high-energy events were identified.
A systematic analysis was conducted to determine the solar context of the events, including the identification of associated X-ray flares and CMEs.
The following solar signatures were searched for, along with the instruments or spacecraft used to observe them:
\begin{itemize}
    \item White-light CMEs – Large Angle and Spectrometric Coronagraph \citep[LASCO;][]{Brueckner1995} on board SOHO;
    \item Type II, III and IV radio bursts – Radio and Plasma Wave instruments (WAVES) on board Wind \citep{Bougeret1995} and the Solar-Terrestrial Relations Observatory \citep[STEREO;][]{Kaiser_2008};
    \item Soft X-ray flares – X-ray Sensor \citep[XRS;][]{Hanser_1996} on board the Geostationary Operational Environmental Satellites (GOES);
    \item Hard X-ray flares – Spectrometer/Telescope for Imaging X-rays \citep[STIX;][]{Krucker_2020} on Solar Orbiter \citep[SO;][]{Muller_2013} and the Reuven Ramaty High Energy Solar Spectroscopic Imager \citep[RHESSI;][]{Lin_2002};
    \item Gamma-ray emission – Large Area Telescope instrument \citep[LAT;][]{Atwood_2009} on board the Fermi Gamma-ray Space Telescope.
\end{itemize}

In addition, information on associated ground-level enhancements \citep[GLEs;][]{Papaioannou_2025} or sub-GLEs \citep{Poluianov_2017} was included when available from the neutron monitor network\footnote{\url{https://gle.oulu.fi/}} \citep{Mavromichalaki_2011}.
The resulting catalogue of these $>$100 MeV proton events associated with their solar signatures is publicly available \citep{Jarry_2025}, and aims to provide a resource for future studies on the origin, acceleration, and propagation of high-energy solar particles.
In this article, we present this catalogue and outline the first results of its analysis.

Section \ref{sec: identification_and_characterization_sep_events} provides details on the data and instrumentation used, as well as the methodologies applied to identify the SEP events.
Then, Sect.~\ref{sec: solar_associations} describes how solar associations were made, including instruments and methods used, as well as some general statistics on the sample.
Section \ref{sec: results} offers an in-depth analysis of the catalogue and provides first results.
Section \ref{sec: conclusion} concludes this work and gives instructions on how to access the catalogue.

\section{SEP event identification and characterisation}
\label{sec: identification_and_characterization_sep_events}

This section outlines the procedure used to identify and characterise the high-energy proton events included in our catalogue. SOHO/ERNE provides a counter for particles penetrating the whole stack, which has not been used before for scientific studies. We adopt it as the baseline dataset since it offers good counting statistics at higher energies and thus useful diagnostic potential.
A systematic scan of SOHO/ERNE/HED penetrating-particle count rates from May 1996 to August 2024 was performed to detect events and determine their onset times.
The corresponding peak fluxes and fluences were derived from SOHO/EPHIN measurements, selected for their long-term continuity and reliability.
Recent recalibration efforts extending EPHIN’s energy range beyond 1~GeV \citep{Kuehl-etal-2017, Kuhl_2019} now provide reliable energy-resolved proton spectra from 100 to about 500~MeV, enabling quantitative flux estimates instead of raw count rates.
In contrast, the ERNE penetrating-particle counter, although sensitive to protons above 100~MeV, still lacks a fully validated response function, preventing direct flux conversion and motivating the use of EPHIN data for flux and fluence calculations.
The methods described below concern only the derivation of catalogue quantities, while the datasets used for solar associations are presented in Sect.~\ref{sec: solar_associations}.

\subsection{Event detection and onset-time determination}
\label{subsec: event_detection}

The events included in the catalogue were identified in the counter data from SOHO/ERNE/HED, specifically using the anti-coincidence channel AC1.
The full time series available which extends from May 1996 to August 2024 were analysed.
For details on the penetrating-particle counter of SOHO/ERNE, see Appendix~\ref{annex: soho-erne_PCC}.
Preliminary detections were obtained using the Poisson–CUSUM method \citep[e.g.][]{Huttunen-heikinmaa_2005, Palmroos_2022}, which identifies statistically significant deviations from a Poisson-distributed background characterised by its mean, $\mu$, and variance, $\sigma$.
When the cumulative sum of deviations remains persistently above a given threshold level, the algorithm flags a potential event.
False or premature detections may occur if the background is noisy, slowly drifting, or poorly estimated.
An SEP event typically exhibits an energy spectrum that decreases steeply with energy, implying higher count rates in detectors sensitive to lower energies. 
Within the SOHO/ERNE/HED telescope, the D3 scintillator layer (see Appendix~\ref{annex: soho-erne_PCC}) registers particles before they reach the anti-coincidence detector AC1.
Therefore, to confirm the detections obtained in AC1, each event was cross-checked in D3, requiring a corresponding enhancement with a higher peak rate than in AC1.
Further details on the algorithm and an example of detection are shown in Appendix~\ref{annex: soho-erne_PCC}.

\begin{figure}[t]
    \centering
    \includegraphics[width=0.99\linewidth]{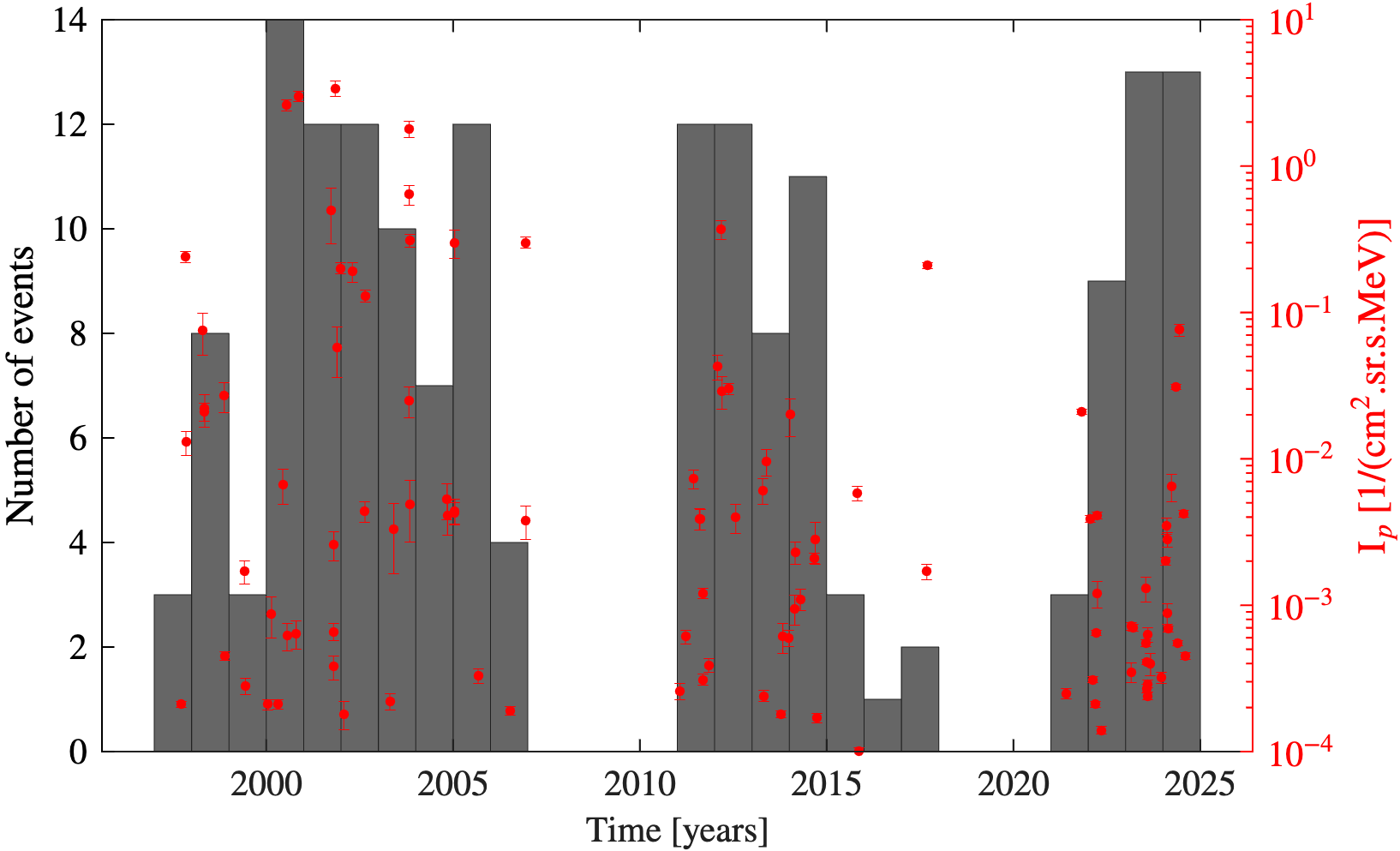}
    \caption{Temporal distribution of the $>$100~MeV proton events included in the catalogue (histogram). Red dots show the corresponding peak proton intensities ($I_\mathrm{p}$) with uncertainties measured by SOHO/EPHIN, plotted against the secondary (right-hand) axis.}
    \label{fig:histo_catalogue}
\end{figure}

The technical adaptation of the aforementioned algorithm was built with the Python package PyOnset \citep{Palmroos2025}.
Occasional false positives were found, often caused by depressed proton counts returning to normal baseline levels \citep[e.g. Forbush decreases; ][]{Papaioannou_2020, Dumbovic_2024}, or detect duplicate events separated by a few data points.
All such cases were manually inspected and corrected.
During the early years of the SOHO mission, several long data gaps occurred in ERNE observations due to instrument or spacecraft anomalies.
Gaps longer than $\sim$18~h between 1996 and 2017 are listed by \citet{Paassilta_2017}, while those exceeding seven days are specified in Appendix~\ref{annex: soho-erne_PCC}.
These interruptions sometimes compromise onset-time accuracy.
No attempt was made to fill these gaps in the present release, leading to the omission of some events (e.g. GLEs~58, 60, and~61) known to involve $>$100~MeV protons.
After the initial scan, approximately 50 obvious false positives were eliminated.
They were mainly due to background rebounds during the decay of previous events or post–Forbush fluctuations.
Further investigation of some of them led to the elimination of six others, resulting in a final list of 172 confirmed events.
These 172 events constitute the $>$100~MeV proton event catalogue.
For each event, the SOHO/ERNE onset time and background mean were recorded, and a unique identifier was assigned following the format \texttt{yyyymmddTHHMMSS}.
In 16 cases where data gaps or very slow rises prevented reliable onset estimation, the identifier was assigned as \texttt{yyyymmddTXXXXXX} to indicate the lack of a precise onset time.
Figure~\ref{fig:histo_catalogue} shows the number of detected $>$100~MeV proton events as a function of time, illustrating the catalogue coverage across solar cycles (SC)~23, 24, and the ongoing SC25. 
A notable feature is the scarcity of such events at the beginning of SC24.
The red scatter points indicate the peak proton intensities ($I_\mathrm{p}$) of these events derived from SOHO/EPHIN measurements.

\subsection{SEP peak flux and fluence determination}
\label{subsec: sep_peak_fluence_determination}

The SOHO/EPHIN instrument measures particles (protons, electrons, and helium nuclei) across a wide energy range, from a few mega-electronvolts up to several hundred mega-electronvolts.
It provides consistent peak flux and fluence estimates as long as only count rates of the previously unused ERNE penetrating-particle counter are available.
A detailed description of the SOHO/EPHIN high-energy proton measurements is given in Appendix~\ref{annex: soho-ephin_instrum}.
Note that we strictly speaking provide only a $>100$~MeV channel proxy that is described as a differential channel.

Mathematically, the integral flux of energetic particles above a given threshold energy, $E_{\mathrm{th}}$, in a SEP event, here $E_{\mathrm{th}}=100$~MeV, is defined as the energy integral of the differential particle flux over all energies exceeding this threshold. If $j(E)$ denotes the differential flux (e.g. in units of particles per square centimetre per second per steradian per mega-electronvolts), the corresponding integral flux $J(>E_{\mathrm{th}})$ is given by
\begin{equation*}
J(>E_{\mathrm{th}}) = \int_{E_{\mathrm{th}}}^{E_{\max}} j(E)\,\mathrm{d}E,
\end{equation*}

where $E_{th}$ and $E_{\max}$ denote the threshold and maximum energy considered, respectively. While $E_{th}=100$~MeV, $E_{max}$ is based on the maximum proton energy obtained during each event. Note that for most GLEs the energy is above the maximum energy of EPHIN ($E_{EPHIN}\approx 500$~MeV) can measure.

In order to be consistent with the ERNE observations, the present analysis derives the integral flux only from a single broad energy channel rather than from multiple discrete energy bins. To characterise the effective energy response of this channel and to determine a representative mean energy, the bow-tie method is applied (for details see appendix \ref{annex: soho-ephin_instrum}). This approach combines the instrument response with assumed spectral shapes, here power laws with indices $\gamma \in \{-5, -2\}$ to define an effective energy and an associated geometric factor (see equation \ref{eq:eq-sulivan} in the appendix \ref{annex: soho-ephin_instrum}). The resulting quantity is a differential channel and is therefore only a proxy for the integral particle flux with energies above $E>100$~MeV. The uncertainty of the EPHIN measurements can be divided into statistical and systematic uncertainties. The systematic uncertainty is estimated to be of the order of 20\%. The statistical uncertainties are determined by the Poisson statistic of the number of selected counts available that are measured in the energy loss interval of interest and vary between a few tens to about 1000 per hour (see Fig.~\ref{fig:EPHIN-Energy-Resolution}). Note that, with respect to the on-board data processing: the statistical error for large events can be larger than that for small events. However, for SEP events, additional factors have to be taken into account for the systematic error, so that in what follows, only the statistical error is provided. 
Note, the derived proxy fluxes are not directly comparable with energetic particle measurements from the GOES satellites, which are provided primarily as integral flux channels above a certain energy threshold (e.g. $>100$MeV). For details, see \citet{Bruno_2017}, \citet{Kress-etal-2021} and references therein.

The $>$100 MeV proton flux derived from EPHIN measurements is averaged over one-hour intervals to ensure sufficient counting statistics. 
Onset times are taken from ERNE measurements, and the same background intervals as those used for ERNE are applied. 
The peak flux is defined as the maximum hourly value within the first few hours after the onset. 
The fluence is computed by integrating the particle flux from the onset until it decreases to 50–10\% of the peak value or falls below $3\sigma$–$5\sigma$ above the background level.
A figure illustrating this procedure is shown in Appendix~\ref{annex: soho-ephin_instrum}.

\begin{figure}[t]
    \centering
    \includegraphics[width=0.9\linewidth]{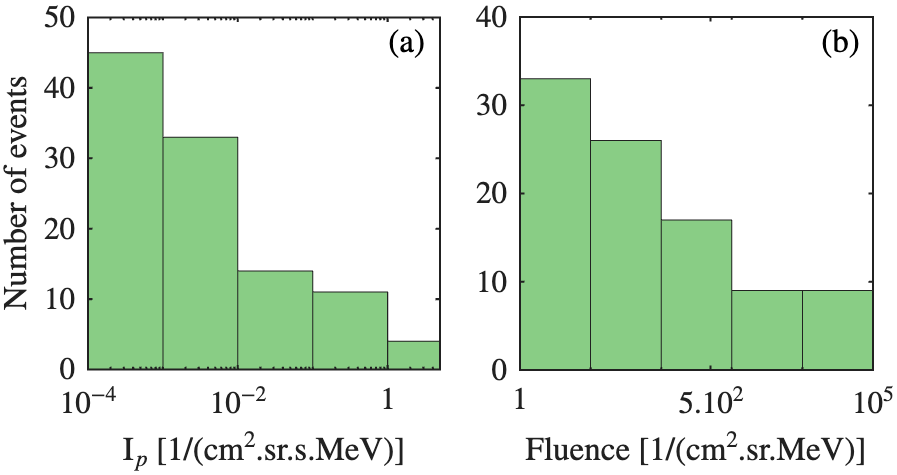}
    \caption{Distributions of $>$100~MeV proton peak intensities (panel~(a)) and fluences (panel~(b)) measured by SOHO/EPHIN for 107 and 97 events of the catalogue, respectively.}
    \label{fig:ephin-peakint-fluence}
\end{figure}

A so-called low-energy-loss statistics (LELS) issue occurs when the EPHIN energy-loss buffer becomes dominated by low-energy particles, leaving too few counts at higher energies (e.g. $>$50~MeV protons) for reliable flux estimation. 
The priority scheme was modified in 2022, which has improved the high-energy statistics since then. 
Another limitation arises during SOHO keyhole periods (KHPs) — intervals typically lasting from a few days to two weeks, when telemetry is reduced or interrupted due to antenna pointing constraints. 
These periods occur once or twice a year at predictable times, determined by the spacecraft’s orbital geometry relative to Earth and the Sun, and can lead to data gaps or degraded measurements.

Figure~\ref{fig:EPHIN-low-statistics} in Appendix~\ref{annex: soho-ephin_instrum} shows two examples of data limitations affecting the $>$100~MeV proton proxy from SOHO/EPHIN.
The published dataset provides, for each event, the background level, peak flux, fluence, and the end date used for the fluence integration. 
Events affected by LELS or keyhole periods (respectively, 53 and 18), are indicated with the labels ‘lels' and ‘khp' in the dedicated ‘comments' column. 
Figure~\ref{fig:ephin-peakint-fluence} summarises the catalogue statistics for the SOHO/EPHIN peak intensities and fluences.

\section{Solar associations}
\label{sec: solar_associations}

The electromagnetic characteristics of the candidate solar sources provide key information on SEP acceleration and transport from the Sun to the point of in-situ measurement. 
Solar associations of the identified SEP events were established through inspection of all available observations, and cross-checked against published catalogues: (1) GOES $>$10~MeV protons\footnote{\url{https://cdaw.gsfc.nasa.gov/CME_list/sepe/}}, (2) \citet{Cane_2010}, (3) \citet{Richardson_2014}, and (4) \citet{Paassilta_2017}.
For events with no ERNE or EPHIN onset determined, onsets were searched in other high-energy datasets, such as STEREO/In-situ Measurements of Particles and CME Transients (IMPACT)/High Energy Telescope (HET) and the GOES/Space Environment Monitor (SEM)/Energetic Particles Sensor (EPS) data (from SEPEM dataset).
When required, lower-energy channels were examined, bearing in mind their slower particle propagation.

\begin{figure}[t]
    \centering
    \includegraphics[width=1\linewidth]{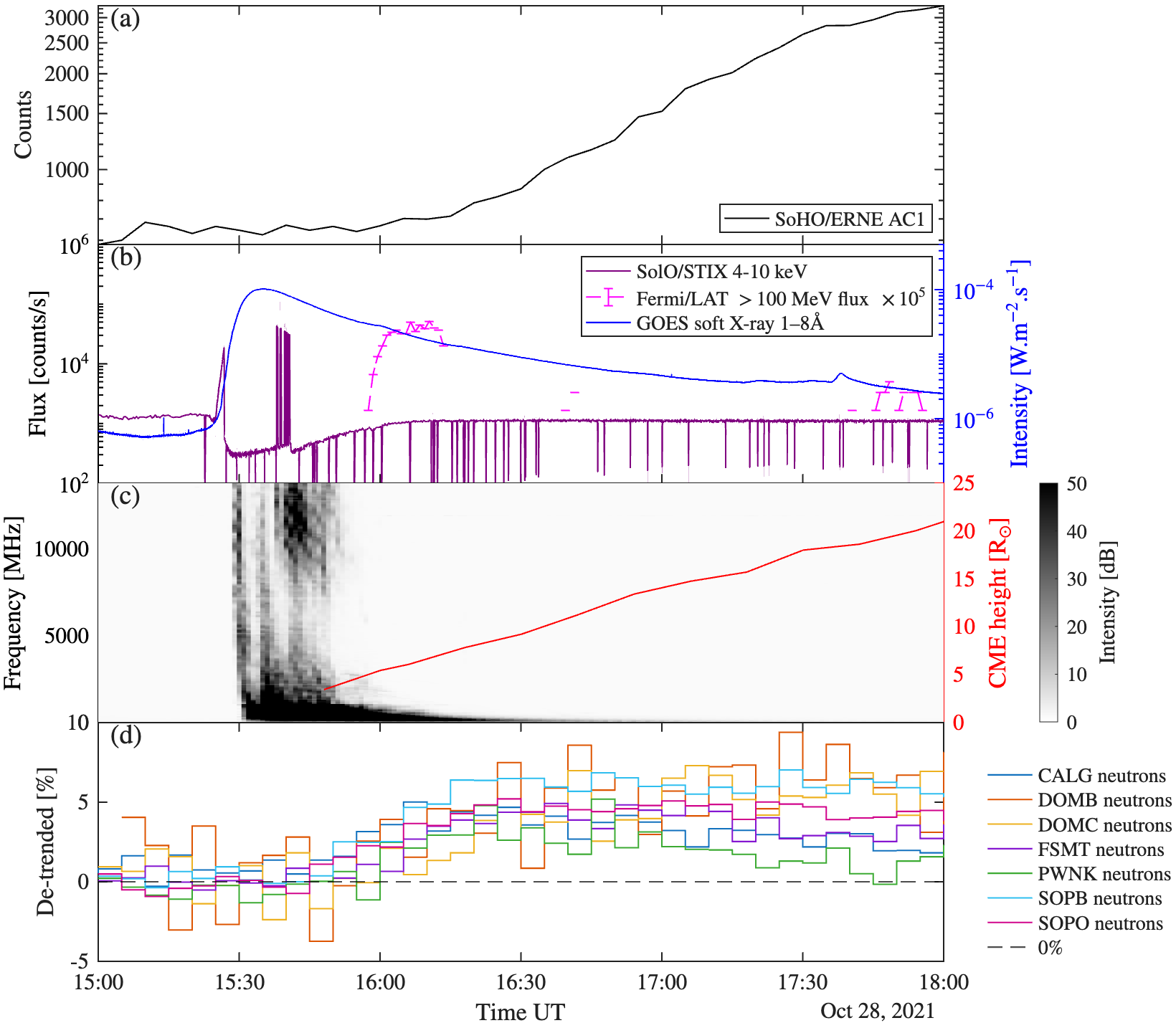}
    \caption{Representative example of the solar associations for the 28 October 2021 $>$100~MeV SEP event (No 137) corresponding to GLE73. 
    Panel (a): ERNE proton counts at $>$100~MeV. 
    Panel (b): Soft X-ray flux (1–8~\AA, GOES, blue) and high-energy flare emissions, including hard X-rays (Solar Orbiter/STIX, 4–10~keV, purple) and $\gamma$-rays (Fermi/LAT, $>$100~MeV flux, pink). 
    Panel (c): Radio dynamic spectrum (e.g. Wind/WAVES) with over-plotted CME height–time profile from coronagraph observations (SOHO/LASCO). 
    Panel (d): Ground-level neutron monitor data showing relativistic particle increases.}
    \label{fig:solar_association_example}
\end{figure}

Only one event showed neither a CME nor a soft X-ray flare; all others exhibited at least one of these solar signatures. 
In total, 73\% of confirmed SEP events (125/172) have clear associations with both a flare and a CME. 
For 22 events, multiple potential solar sources were identified; the selected one is reported in the catalogue, while alternatives and differences with published catalogues are noted in the ‘general notes' (Appendix~\ref{annex: structure_catalogue}).
Figure~\ref{fig:solar_association_example} illustrates, for one representative $>$100~MeV SEP event, the various solar signatures considered in this work — proton flux, flare emissions, radio and CME signatures, and GLEs.
Dates in the catalogue follow the ISO 8601 format (yyyy-MM-ddTHH:mm:ssZ), where the trailing Z indicates UTC time.
Each group of columns includes a comments field providing additional relevant information.
The detailed catalogue structure is described in Appendix \ref{annex: structure_catalogue}.

\subsection{Identification of soft X-ray flares}
\label{subsec: cat_SXR}

Soft X-ray (SXR) enhancements are observed during solar flares, one of the two main accelerators of SEPs.  
SXR flare intensity correlates with CME speed and shock strength \citep{Salas-Matamoros_2015, Jarry_2023, Papaioannou_2024}, and with SEP event intensity \citep[e.g.][]{Papaioannou_2016, Kouloumvakos_2019}.  
Strong flares are typically linked to fast, wide CMEs and intense SEP events.

\begin{figure}[t]
    \centering
    \includegraphics[width=0.9\linewidth]{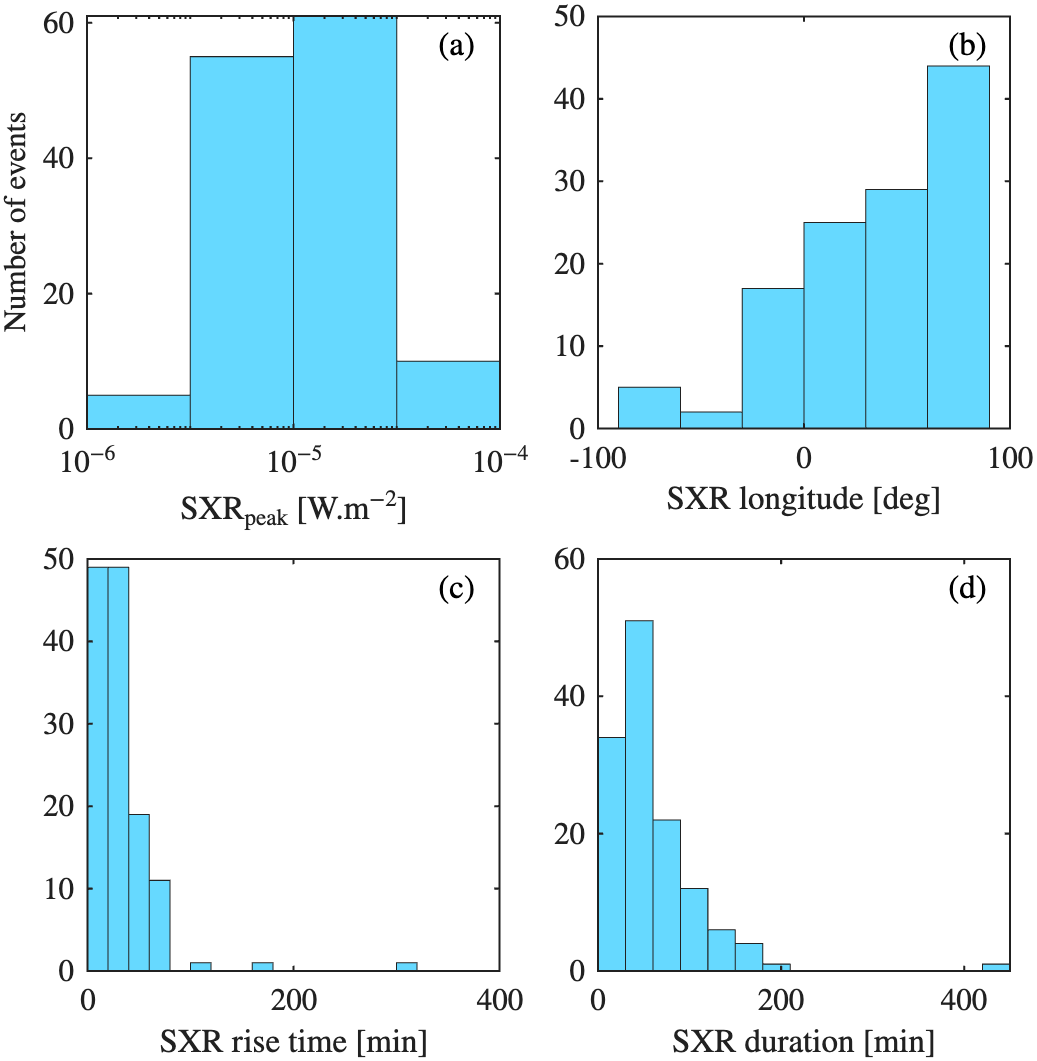}
    \caption{Statistics of the identified soft X-ray flares associated with $>$100~MeV SEP events (131/172, 76\%).  
    From left to right: SXR peak flux, flare longitude in Stonyhurst co-ordinates, rise time (peak minus start time), and total duration.}
    \label{fig:sxr_stats}
\end{figure}

To identify SXR flare associations, the solar particle release (SPR) time was estimated using the IRAP Propagation Tool \citep{Rouillard_2017} in ‘SEP propagation mode'.  
Assuming scatter-free propagation along the Parker spiral, the tool provides the solar footpoint and SPR time based on in-situ solar wind speed.  
Candidate flares were then searched around or before this time within the active-region belt.
SXR fluxes were obtained from the GOES X-ray Sensor (XRS)\footnote{\url{https://www.ngdc.noaa.gov/stp/satellite/goes-r.html}}, operating since 1975 under NOAA \citep{Woods_2024}.  
The 1–8~\AA\ channel defines the standard flare classification (A–X classes for fluxes of 10$^{n}$~W~m$^{-2}$, $n$=[-8...-4]). 
Following \citet{Hudson_2024} and the reprocessed science-quality data from NOAA National Centers for Environmental Information (NCEI), a calibration factor of 1.43 was applied to the GOES-1 through GOES-15 SXR datasets.
Data cadence ranges from 1~min (GOES 8–12) to 1–2~s (GOES 13–18).
For each SEP event, the most intense flare within the derived window was selected.  
Ambiguous cases with multiple nearby flares were reviewed individually, and alternatives are noted in the catalogue comments.  
Events with flares located behind the solar limb were marked as ‘far side' and excluded from flare analysis (41/172 events).  
An example of GOES SXR flux is shown in panel~(b) of Figure~\ref{fig:solar_association_example}.
Flare start, peak, and end times, as well as H$\alpha$ locations, were taken from the GOES catalogues\footnote{1) \url{https://www.ngdc.noaa.gov/stp/space-weather/solar-data/solar-features/solar-flares/x-rays/goes/xrs/} (pre-Jun 2017), 2) \url{https://umbra.nascom.nasa.gov/goes/eventlists/goes_event_listings/}}.  
For events after October 2002, locations were verified and completed using Solar Dynamics Observatory (SDO)/Atmospheric
Imaging Assembly (AIA) EUV observations from SolarSoft\footnote{\url{https://www.lmsal.com/solarsoft/latest_events_archive.html}}.  
Discrepancies between catalogues occurred in only three cases (Events 103, 113, and 120), for which SolarSoft values were adopted.  
One remaining case (Event~18) was completed using \citet{Cane_2010}.  
SXR peak fluxes were extracted from GOES time series\footnote{\url{https://www.ncei.noaa.gov/products/goes-r-extreme-ultraviolet-xray-irradiance}}.  
Fluences were obtained by trapezoidal integration between the reported start and end times.  
The catalogue specifies the GOES spacecraft used (‘Observer' column) and provides SolarSoft links and positional remarks in the comments (‘behind the limb' or ‘far side'). Overall, 131 of 172 SEP events (76\%) were clearly associated with an SXR flare, while 41 were far-side events.
Figure~\ref{fig:sxr_stats} shows the distribution of flare peak intensities, longitudes, rise times, and durations.  
Flare longitudes range from $-89^{\circ}$ to $132^{\circ}$, with a mean (median) of $49.9^{\circ}$ ($54.5^{\circ}$),  whereas latitudes range from $-37^{\circ}$ to $30^{\circ}$, with a mean (median) of $-1.2^{\circ}$ ($-4^{\circ}$).
The predominance of western flares, and the low dispersion of latitudes around the equator, indicates the importance of an optimal magnetic connection of the source to Earth for $>$100~MeV detections.  
SXR peak fluxes span $2.5\times10^{-6}$–$2.8\times10^{-3}$~W~m$^{-2}$, with a mean of $2.8\times10^{-4}$~W~m$^{-2}$ and a median of $1.1\times10^{-4}$~W~m$^{-2}$, confirming that strong SEP events are primarily associated with strong ($>$M-class) flares.

\subsection{Identification of hard X-rays}
\label{subsec: cat_HXR}

\begin{figure}[t]
    \centering
    \includegraphics[width=0.9\linewidth]{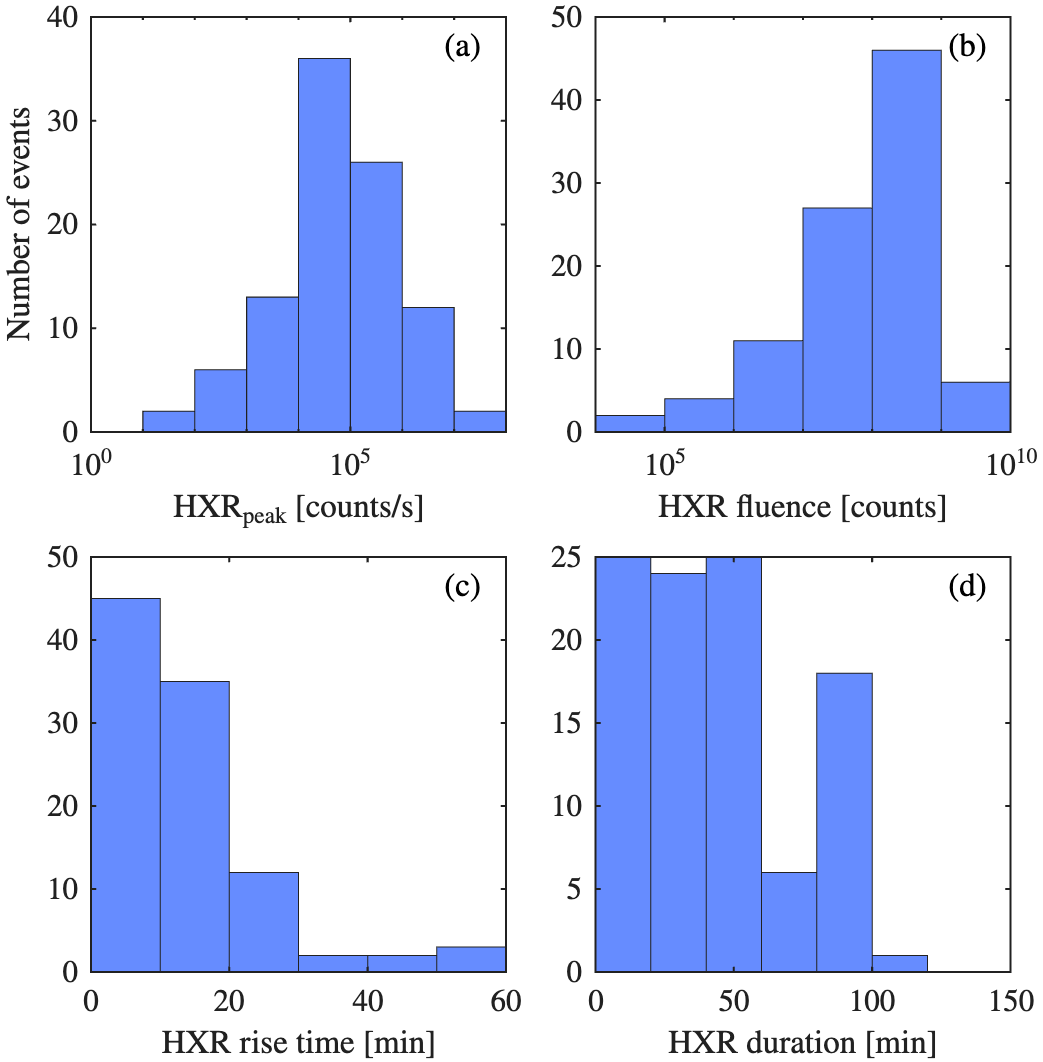}
    \caption{Statistics of the identified hard X-ray flaring sources for the $>$100~MeV catalogue (76\%, 131/172 events).
    From left to right: distribution of HXR peak intensity, flare fluence (in counts), rise time (HXR peak -- start), and total duration.}
    \label{fig:hxr_stats}
\end{figure}

Hard X-ray (HXR) observations provide key diagnostics of flare-accelerated electrons and the hottest plasma components. 
Unlike SXRs, mainly produced by $\sim$10–30~MK thermal plasma (see Sect.~\ref{subsec: cat_SXR}), HXRs are dominated by bremsstrahlung emission from non-thermal electrons, i.e from energetic electrons colliding with ambient protons, with a thermal contribution at the lower energies from plasma hotter than $\sim$30~MK. 
The strongest signatures of particle acceleration appear during the impulsive phase, typically as bursts of HXR and microwave emission, both tracing non-thermal electron populations \citep[see e.g.][]{Klein_2021}.

HXR flare associations were established from two complementary instruments. 
RHESSI (2002–2018) provided HXR and $\gamma$-ray measurements from 3~keV to 20~MeV, while the Spectrometer Telescope for Imaging X-rays (STIX) on board Solar Orbiter (SO) has operated since 2020, offering HXR imaging spectroscopy between 4–150~keV. For events~43–134, both RHESSI time series (retrieved from the RHESSI data browser\footnote{\url{https://hesperia.gsfc.nasa.gov/rhessi3/}}) and the RHESSI flare catalogue\footnote{\url{https://hesperia.gsfc.nasa.gov/hessidata/dbase/hessi_flare_list.txt}} provided by NASA were used. 
This catalogue lists detections in the 6–12~keV range, providing start, end, and peak times, as well as peak and total counts. 
For events~135–172, analysis relied on the SO/STIX time series from the STIX data browser\footnote{\url{https://datacenter.stix.i4ds.net/view/plot/lightcurves\#}} and its science flare list\footnote{\url{https://datacenter.stix.i4ds.net/view/flares/list}}, based on detections in the 4–10~keV band. 
Note that RHESSI and STIX counts correspond to different energy ranges and response matrices; therefore, peak intensities are not directly comparable across instruments.
Associations between $>$100~MeV SEP events and HXR flares were established by comparing flare timing in hard and soft X-rays. 
Only HXR flares peaking after the SXR onset ($\pm$1~min) and no later than 30~min after the SXR peak were retained, consistent with the Neupert effect \citep{Neupert_1968}, which links impulsive HXR emission to the subsequent rise of SXR emission. 
A few exceptions were made: 
for SEP event~50 (22~August~2002), the nearest HXR flare was spatially distinct from the SXR source, so the preceding co-located flare was selected instead; 
for event~152 (17~July~2023), the HXR peak occurred 10~min before the SXR onset but overlapped in duration and was particularly intense; 
for events~81 (7~September~2005) and~119 (7~January~2014), HXR peaks occurred 40–42~min after their SXR peaks. 
Solar Orbiter catalogues, namely the Solar Orbiter's Comprehensive Solar Energetic Electron event Catalogue \citep[CoSEE-Cat;][]{Warmuth_2025} and the Solar Orbiter large SEP events catalogue (Papaioannou et al., \textit{in prep.}), were also used to cross-check STIX associations. 
Because RHESSI and STIX provide limited coverage restricted to their operational periods and visible disc, occulted flares may be only partially or not at all detected.
For each event, we list the start, peak, and end times, peak and total counts, maximum energy range, flare location in Carrington co-ordinates, and the observing spacecraft.
A comment column indicates whether the flare was on the far side or behind the limb. Far-side events were not observed, while behind-the-limb events were measured but not positioned, and possibly occulted.
Among the events for which an HXR search was possible (No.~43–172, as the first 42 precede RHESSI), 19\% (25/130) were far side and 4\% (5/130) lacked RHESSI data.
Figure~\ref{fig:hxr_stats} summarises the measured HXR peak intensities, fluences, rise times, and durations.
The mean HXR magnitude is $8.2\times10^{5}$~counts, ranging from 23 to $1.6\times10^{7}$~counts, with a median of $5.9\times10^{4}$~counts.

\subsection{Identification of type III radio bursts}
\label{subsec: cat_radiotIII}

Radio bursts are classified into five categories (types~I–V) based on their frequency–time spectrograms \citep{Wild_1963}.
Type~III bursts are the most common coherent solar radio emissions, produced by beams of flare-accelerated electrons (tens to hundreds of~kilo-electronvolts) escaping along open magnetic field lines into the interplanetary (IP) medium \citep{Reid_2017}.
Their rapid frequency drift reflects electron propagation through progressively less dense plasma, generating emission at the local plasma frequency and its harmonic.
They often occur in groups during major solar eruptions, indicating successive magnetic reconnection episodes (see panel~(c) of Fig.~\ref{fig:solar_association_example}).
Using coronal density models, the timing of a type~III burst can therefore constrain the release time and coronal height of the associated SEP event.

The search for type~III associations combined space- and ground-based radio observations to ensure full spectral and temporal coverage.
Ground-based instruments typically observe from $\sim$10~MHz to a few~gigahertz, while lower frequencies are blocked by the ionosphere.
We used data from observatories worldwide covering 10~MHz–2.5~GHz, including stations in Australia, the USA, Japan, Greece, and France.
Space-based receivers, Wind/WAVES and the twin STEREO/WAVES instruments \citep{Bougeret2008}, extend the coverage to kilohertz–megahertz frequencies originating in the outer corona and IP medium.
All instruments used for identifying radio signatures of the SEP events are summarised in Appendix~\ref{annex: radio_instruments}.
Type~III bursts were identified by their temporal association with the start and peak times of the parent SXR flare.
For far-side events, the timing was instead compared with the CME onset or any corresponding type~II burst (see Section~\ref{subsec: cat_radiotII}).
Spectra from WIND/WAVES\footnote{\url{https://spdf.gsfc.nasa.gov/pub/data/wind/waves/}} and STEREO/WAVES\footnote{\url{https://stereo.space.umn.edu/data/level-3/STEREO/Both/SWAVES/daily-summary-plots/}} were visually inspected to determine the associated type~III start time and frequency range, and this informations was reported in the catalogue along with the observing instrument(s) (SWAVES and/or WAVES).
Ground-based identifications were obtained from compiled NOAA\footnote{\url{https://www.ngdc.noaa.gov/stp/space-weather/solar-data/solar-features/solar-radio/radio-bursts/reports/spectral-listings/}} and SolarMonitor\footnote{\url{https://www.solarmonitor.org/data/}} reports, which include observations from the Radio Solar Telescope Network (RSTN) and other stations.
When available, event lists from the e-Callisto network\footnote{\url{https://www.e-callisto.org/}} and composite dynamic spectra from the LESIA Radio Monitoring service\footnote{\url{https://secchirh.obspm.fr/spip.php?article11}} were also examined.
For ground-based detections, we recorded the observing station(s), start time, and full frequency range; when multiple stations reported the event, the earliest start time and the minimum–maximum frequency limits across all reports were adopted.
Type~III radio bursts were identified for all (100\%) SEP events in the catalogue, although some were detected only from space or from the ground.
All associations were cross-checked with published catalogues, and matching identifications are noted in the type~III comments.
The referenced catalogues are \citet{Papaioannou_2016}, covering 1998–2013, and the PhD thesis of A.~Kouloumvakos (2017)\footnote{\url{https://www.didaktorika.gr/eadd/handle/10442/44779}}, spanning 1997–2010.

\subsection{Identification of type II radio bursts}
\label{subsec: cat_radiotII}

Type~II radio bursts trace shock waves accelerating particles in the solar atmosphere, typically driven by fast CMEs whose speeds exceed the local Alfvén speed \citep[][and references therein]{Kumari_2023}, and less frequently by termination shocks from powerful flares \citep{Hillaris_2006}.
Such shocks are strong candidates for producing very high-energy SEPs \citep{Klein_2021b}.
In dynamic spectra, type~II bursts appear as slowly drifting, narrowband emissions reflecting the decreasing plasma density as the shock moves outwards. 
They are usually identified by visual inspection since they often occur close in time to other emissions (type~III or IV). Ground-based observations commonly show a rapidly drifting type~III burst (flare electrons) followed by the slower type~II emission (shock vicinity), while space-based measurements usually provide a clearer separation as electron beams propagate faster than the shock. 
Morphological characteristics depend on the shock strength and ambient plasma conditions \citep{Wild_1963}, making type~II bursts valuable diagnostics of shock dynamics and particle acceleration regions tracing.

Type~II associations for SEP events were retrieved from the same space- and ground-based instruments used for type~III identifications (Sect.~\ref{subsec: cat_radiotIII}), ensuring broad spectral and temporal coverage of this study to trace shocks from the low corona to the IP medium. 
Associations were based on CME release times.
Type~II bursts were identified by visual inspection using standard criteria based on their slow drift rate (as compared to type III bursts), harmonic structure (fundamental and second harmonic emission lanes), and morphology which sometimes exhibits fragmented or ‘herringbone' structure. Quantitative thresholds, for example on minimum burst duration or signal-to-noise ratio were not applied.
For each identified type~II burst in space-borne spectra we recorded the observing instrument (SWAVES and/or WAVES), start time, and frequency range. 
Additional information was taken from the multi-vantage-point type~II catalogue\footnote{\url{https://cdaw.gsfc.nasa.gov/CME_list/radio/multimission_type2/Full_catalogue.html}} of \citet{Mohan2024}. 
Ground-based detections were catalogued with observing station(s), start time, and frequency range; when multiple stations reported the event, we recorded all of them, the earliest start time, and the minimum–maximum frequencies observed.
Positive type~II identifications were established for most SEP events (163/172, i.e. 95\%), cross-checked with several published catalogues: the continuously updated \citet{Gopalswamy_2019}\footnote{\url{https://cdaw.gsfc.nasa.gov/CME_list/radio/waves_type2.html}} list (1996–2024), \citet{Papaioannou_2016} (1998–2013), and the PhD thesis of A.~Kouloumvakos (2017)\footnote{\url{https://www.didaktorika.gr/eadd/handle/10442/44779}} covering 1997–2010.

\subsection{Identification of type IV radio bursts}
\label{subsec: cat_radiotIV}

Type~IV radio bursts appear as broadband continuum emissions in dynamic spectra, generally linked to major flares or CMEs \citep{Kahler_1992}. 
They are classified as moving or stationary depending on whether a clear frequency drift is observed.
In the context of SEP events, type~IV bursts trace prolonged electron acceleration and trapping in coronal magnetic structures, complementing X-ray and coronagraph data in characterising the timing and nature of acceleration sites \citep{Mohan_2024}.

Space-based type~IV associations were obtained directly from the \citet{Mohan_2024}\footnote{\url{https://cdaw.gsfc.nasa.gov/CME_list/radio/type4/DHtypeIV_catalog.htm}} catalogue, which uses Wind/WAVES and both STEREO/WAVES (Ahead and Behind) observations. 
For each associated event, we list the observing spacecraft, onset time, duration (in minutes), and end frequency (in~megahertz). 
Complementary ground-based detections were retrieved from the sources described in Sect.~\ref{subsec: cat_radiotIII} and Appendix~\ref{annex: radio_instruments}, with the corresponding station(s), start time, and frequency range included in the catalogue.
Type~IV radio bursts were identified for 53\% of the events (91/172). 
Their detection remains challenging due to variable spectral morphologies, frequent overlap with type~II and type~III emissions, and gaps in instrumental frequency coverage. 
These factors likely contribute to the relatively low association rate derived in this study.

\subsection{Identification of $\gamma$-ray emissions}
\label{subsec: cat_gammaray}

Observations of solar $\gamma$-rays and neutrons provide direct diagnostics of high-energy ions accelerated during eruptions. 
These ions interact with the lower solar atmosphere, producing $\gamma$-ray emission through nuclear reactions \citep{Vilmer_2011}. 
Ions in the 1–100~MeV/nuc range generate line emissions from nuclear de-excitation, neutron capture, and positron annihilation, while energies above 100~MeV/nuc are dominated by neutral pion decay.
Solar $\gamma$-ray emission typically exhibits two phases \citep{Frost_1971}: a prompt component coinciding with the flare, and a delayed, prolonged component that persists well after the impulsive flare signatures have faded \citep{Akimov_1991, Ryan_2000, Rank_2001}.
It is generally accepted that the first component is linked to ion acceleration in the reconnection region, whereas the origin of the delayed phase remains unresolved and actively debated.
Proposed scenarios include continuous acceleration or trapping in large coronal structures \citep[e.g.][]{Ryan_2000, Grechnev_2018, deNolfo_2019, Bruno_2023}, as well as acceleration at CME-driven shocks \citep[e.g.][]{Ackermann_2014, Share_2018}.
Durations of sustained $\gamma$-ray events correlate with those of $>$100~MeV SEP events, though the latter are usually about five times longer \citep{Share_2018}.
Fermi/Large Area Telescope (LAT) has observed $\gamma$-ray emission from behind-the-limb flares \citep{Ackermann_2017}, demonstrating that pion-decay $\gamma$-rays can originate tens of degrees from the active region \citep{Plotnikov_2017}.
Although correlations with SEP events and CME properties are often reported, notable counterexamples and strong event-to-event variability indicate that these relationships are not universal, and that sustained $\gamma$-ray emission cannot be explained by a single acceleration mechanism.
The Fermi space telescope and its two main instruments provide systematic detection of these two phases since 2011.
The LAT detects photons from 30~MeV to $>$300~GeV \citep{Lange_2013, Ajello_2021}, while the Gamma-ray Burst Monitor (GBM) covers a few kilo-electronvolts to $\sim$30~MeV \citep{Meegan_2009}, bridging low-energy X-rays and high-energy $\gamma$-rays.

We analysed publicly available GBM\footnote{Fermi/GBM: \url{https://hesperia.gsfc.nasa.gov/fermi/gbm/qlook/}} data for SEP events~87–172 (2011–2024), corresponding to the Fermi mission period, to establish SEP–$\gamma$-ray associations by temporal correspondence with sustained emission periods, with no ambiguous overlaps.
The Fermi/LAT Significant Event List\footnote{\url{https://hesperia.gsfc.nasa.gov/fermi/lat/qlook/}} ($>$100 MeV) and Fermi-LAT Solar Flare Catalog \citep[30 MeV–10 GeV,][]{Ajello_2021b} were also inspected over the same periods.
An example of a Fermi/LAT light curve is shown in panel (b) of Fig. \ref{fig:solar_association_example}.
For each associated event, the catalogue lists from Fermi/GBM the onset, peak, and end times, peak intensity (counts~s$^{-1}$), total counts, and observing instrument, with comments noting missing Fermi data or flare absence.
Among the 87 SEP events occurring during the Fermi era, 65 (75\%) show associated $\gamma$-ray emission. 
Non-detections may result from insufficient proton energies ($>$300~MeV required for pion decay), Fermi’s orbital night-time gaps, or far-side eruptions magnetically disconnected from the visible disc.

\subsection{Identification of coronal mass ejections (CMEs)}
\label{subsec: cat_CME}

Coronal mass ejections are sudden large-scale releases of plasma ejected into the upper corona and the solar wind.
They result from the reconfiguration of coronal magnetic fields through various processes and strong ones often accompany flares or prominence eruptions \citep{Yashiro_2005}. 
Fast and wide CMEs represent key components of major solar storms, typically involving (1) an initiation phase with flaring or prominence activity, (2) coronal expansion possibly driving shocks and large-scale solar atmosphere disturbances, and (3) the ejection of mass and magnetic flux into interplanetary space. 
When a CME appears as a halo-like enhancement around the coronagraph’s occulting disc, it is classified as a halo CME \citep[see][and references therein]{Webb_2012}.

\begin{figure}[t]
    \centering
    \includegraphics[width=0.9\linewidth]{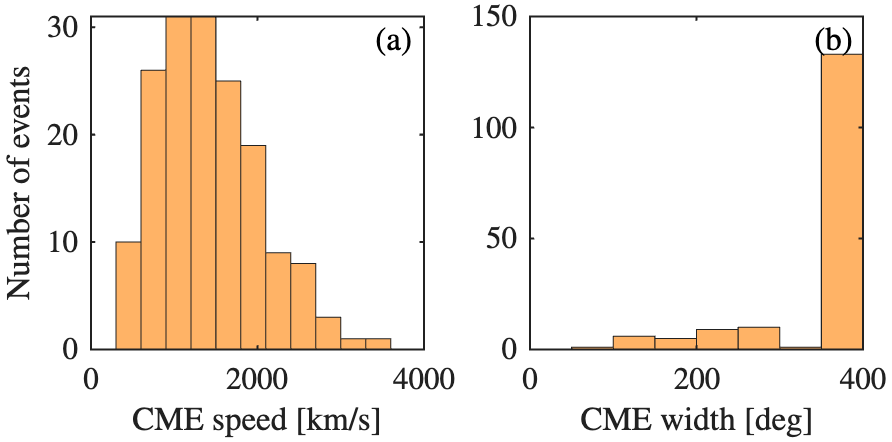}
    \caption{Statistics on the CMEs associated with the $>$100~MeV events (96\%, 162/169 events).
    Panel (a): CME speed distribution (kilometre per second).
    Panel (b): CME width distribution (deg).}
    \label{fig:cmes_speed_and_width}
\end{figure}

Associations between SEP events and CMEs were determined using LASCO C2 and C3 coronagraphs on board SOHO, which have continuously monitored the solar corona since 1996. 
All manually identified CMEs from these instruments are compiled in the SOHO/LASCO CME catalogue \citep{Gopalswamy_2009}, available at the CDAW Data Center\footnote{\url{https://cdaw.gsfc.nasa.gov/CME_list/}}. 
Coronal mass ejection linear speeds were derived from first-order polynomial fits to height–time measurements (see panel (c) of Fig.~\ref{fig:solar_association_example}) listed in that catalogue. 
For each SEP event, CMEs occurring after the onset of the associated SXR flare (Sect.~\ref{subsec: cat_SXR}) were considered; in the absence of recorded SXR data, the SEP onset time was used instead. 
Given the high energies of $>$100~MeV SEP events, their CMEs are generally fast and well defined in LASCO observations, although some events coincide with LASCO data gaps, explicitly flagged in the catalogue.
For each CME, the catalogue lists the first LASCO~C2 appearance time, linear speed, angular width, and a direct link to the corresponding movie in the LASCO database. 
Coronal mass ejections with apparent widths of 360$^{\circ}$ are identified as halos. 
Associations were found for 96\% (165/172) of SEP events; the remaining cases correspond to SOHO/LASCO data gaps, noted as `NO DATA' in the source field. Figure~\ref{fig:cmes_speed_and_width} summarises CME kinematics for $>$100~MeV proton events. 
Coronal mass ejection speeds range from 448 to 3387~km\,s$^{-1}$, with a median of 1336~km\,s$^{-1}$ and a mean of 1404~km\,s$^{-1}$. 
Widths span from 76$^{\circ}$ to 360$^{\circ}$, with mean and median values of 331$^{\circ}$ and 360$^{\circ}$, respectively. 
These statistics confirm that strong SEP events are predominantly associated with fast, halo CMEs \citep[see e.g.][and references therein]{Papaioannou_2016}.

\subsection{Identification of ground-level enhancements (GLEs)}
\label{subsec: cat_GLE}

Ground-Level Enhancements are rare, high-intensity SEP events in which protons reach relativistic energies (up to the GeV range) and penetrate the Earth’s atmosphere, producing secondary particle cascades detected by ground-based neutron monitors worldwide \citep{Gopalswamy_2012, Papaioannou_2023macau}. 
A GLE is defined as a near-simultaneous increase in at least two neutron monitors, including one near sea level, while events detected only at high-altitude stations are termed sub-GLEs \citep{Poluianov_2017}.
Representing the most energetic end of the SEP spectrum over the past seven decades, GLEs provide unique diagnostics of extreme particle acceleration at the Sun \citep{Kocharov_2023}.
Their properties \citep{Asvestari_2017} have been used to analyse extreme SEP events \citep{Usoskin_2023} and to quantify their impact on the solar radiation environment \citep{Herbst_2025}.
GLEs are generally associated with powerful flares and fast CMEs driving strong shocks \citep[see e.g.][and references therein]{Mishev_2021, Mishev_2022, Mishev_2024}, but their occurrence is also favoured by good magnetic connectivity and by the presence or interaction of transient solar-wind structures—such as ICMEs, magnetic clouds, and shocks \citep{Rouillard_2012, Papaioannou_2016, Lario_2017}, making them particularly complex to fully understand \citep{Cliver_2006}.

GLE and sub-GLE associations were identified manually using the neutron monitor database\footnote{\url{https://gle.oulu.fi/}}.
For each event, the catalogue lists the maximum intensity, the observing station, and the GLE identifier (or ‘sub-GLE' label).
When available, fluences from \citet{Usoskin_2020} are also included.
The catalogue contains 18 GLEs and 4 sub-GLEs.
All GLEs and sub-GLEs observed during the studied period are represented, except GLEs~58, 60, and~61, which are absent due to SOHO/ERNE data gaps.
Fluence values were taken by \cite{Usoskin_2020} who derive full-event, de-trended, time-integrated NM intensities that include both the early anisotropic and later isotropic phases.
Thus, unlike first-hour–focused inversions, their reported values represent the total omnidirectional fluence. These are available for 11 of the 22 recorded events.

\section{Results}
\label{sec: results}

Table~\ref{tab:av_values_solar_assoc} summarises the principal statistical properties of the $>$100~MeV events and their solar associations. 
Where data exist, every SEP event is associated with a CME and a SXR flare, implying particle acceleration by the flare, the CME-driven shock, or both. 
Median CME speeds exceed 1300~km\,s$^{-1}$ and median CME angular widths are $\sim360^\circ$ (Table~\ref{tab:av_values_solar_assoc}).

\begin{table}[b]
    \centering
    \caption{Median and mean values with standard deviations ($\sigma$) of the parameters characterising the $>$100~MeV proton events and the associated CMEs, soft and hard X-ray flares, and $\gamma$-ray emissions.}
    \renewcommand{\arraystretch}{1.12}
    \resizebox{\columnwidth}{!}{%
    \begin{tabular}{@{} l c c c @{}}
        \toprule
        Parameter [unit] & $N$ & Median & Mean $\pm \sigma$ \\
        \midrule
        I$_p$ [1/(cm$^2$ sr s MeV)] & 107 & $2.3 \times 10^{-3}$ & $0.14 \pm 0.53$ \\
        Fluence [1/(cm$^2$ sr MeV)] & 97 & 36 & $4.2 \pm 14.5 \times 10^{3}$ \\
        \midrule
        V$_{\mathrm{CME}}$ [km/s] & 164 & 1345 & $1411 \pm 608$ \\
        Angular width [$^{\circ}$] & 165 & 360 & $331.9 \pm 63.7$ \\
        \midrule
        SXR$_{\mathrm{peak}}$ [W/m$^2$] & 131 & $1.1 \times 10^{-4}$ & $2.7 \pm 4.7 \times 10^{-4}$ \\
        SXR fluence [J/m$^2$] & 130 & $1.6 \times 10^{-2}$ & $8.8 \pm 21.9 \times 10^{-2}$ \\
        SXR longitude & 122 & 38 & $35.5 \pm 42.4$ \\
        SXR rise time [min] & 131 & 23 & $32 \pm 33.2$ \\
        SXR duration [min] & 131 & 44 & $59.4 \pm 50.3$ \\
        \midrule
        HXR$_{\mathrm{peak}}$ [counts/s] & 97 & $5.9 \times 10^{4}$ & $8.2 \pm 23.9 \times 10^{5}$ \\
        HXR fluence [counts] & 96 & $1.2 \times 10^{8}$ & $2.5 \pm 4.2 \times 10^{8}$ \\
        HXR rise time [min] & 99 & 10.8 & $13.2 \pm 12.5$ \\
        HXR duration [min] & 99 & 40.6 & $44.6 \pm 29.1$ \\
        \midrule
        $\gamma_{\mathrm{peak}}$ [counts/s] & 63 & $6.8 \times 10^{4}$ & $2.7 \pm 3.5 \times 10^{5}$ \\
        $\gamma$-rays fluence [counts] & 63 & $2.6 \times 10^{7}$ & $1.8 \pm 3.1 \times 10^{8}$ \\
        $\gamma$-rays duration [min] & 64 & 27.5 & $7.3 \pm 177.4$ \\
        \bottomrule
    \end{tabular}
    }
    \label{tab:av_values_solar_assoc}
\end{table}

\begin{figure}[t]
    \centering
    \includegraphics[width=\linewidth]{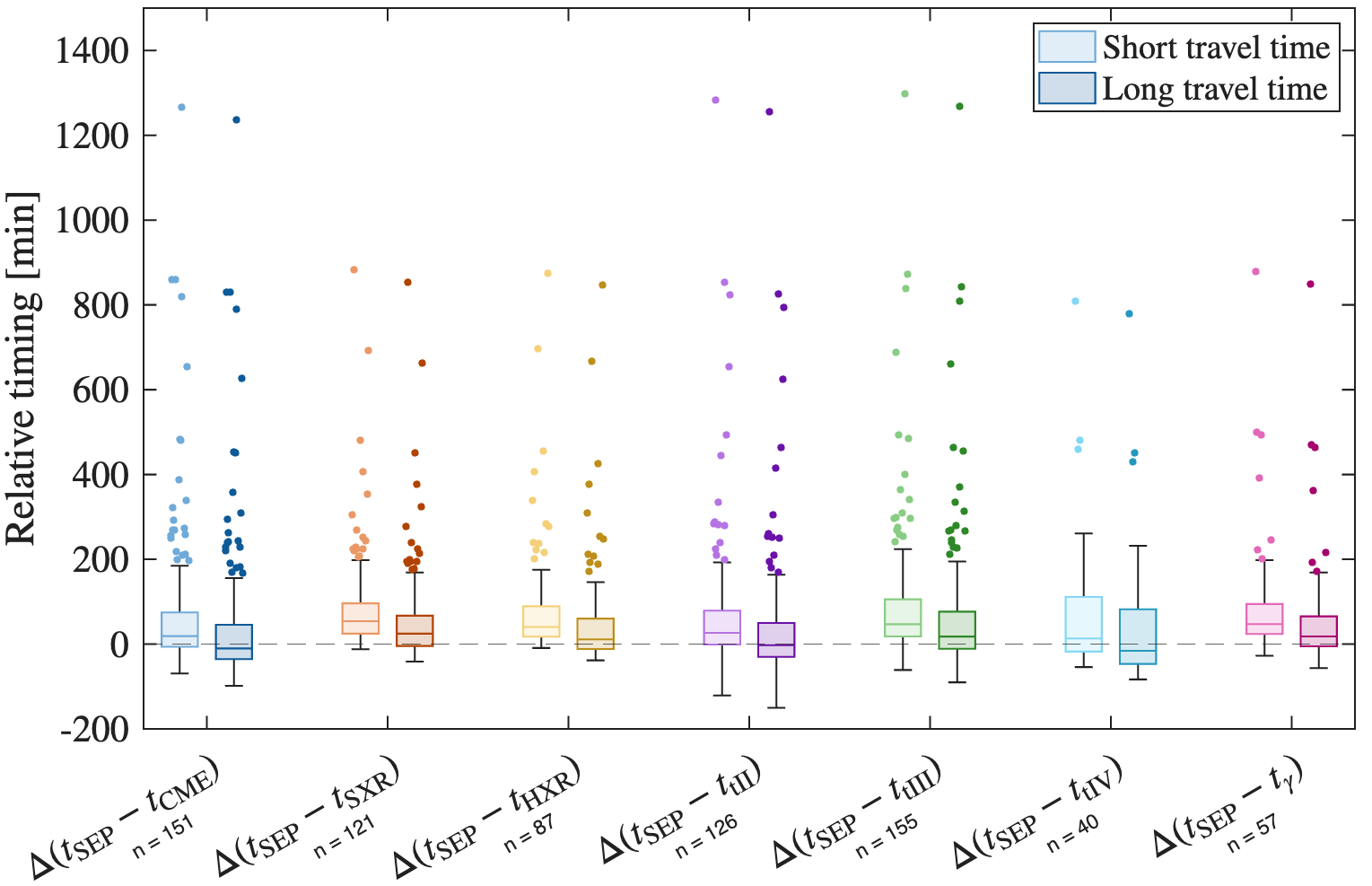}
    \caption{Distribution of the SEP events with respect to timing differences (in minutes) between the $>$100~MeV proton events release times determined with SOHO/ERNE SEP onset and the associated solar signatures onsets.
    Two travel-time assumptions are considered: short (light colours) and long (dark colours) path lengths, respectively, 1 and 2.5 AU.
    Solar signatures are shifted to the solar time according to the light travel time, and are, from left to right: CME first appearance in SOHO/LASCO coronographs (blue), soft X-ray onset time (red), hard X-ray onset time (yellow), type II radio burst onset time (purple), type III radio burst onset time (green), type IV radio burst onset time (cyan), gamma ray burst onset time (pink).
    In each boxplot, the middle line indicates the median of the distribution whereas the top of bottom edges of a box are the upper and lower quartiles, respectively.
    Outliers are values that are more than 1.5 times the distance between the top and bottom edges.
    The number of events considered in each boxplot is indicated in the x axis.}
    \label{fig:sol-asso-SEP_delays}
\end{figure}

The typical travel time of 100~MeV protons is 19–49~min depending on the path length (1–2.5~AU). 
Including light-travel delays, the observed particle onset should therefore occur within roughly one hour of its parent solar signature.
Figure~\ref{fig:sol-asso-SEP_delays} presents timing differences between SEP release times (from ERNE onset) and associated solar signatures (CME, SXR, HXR, type~II/III/IV, $\gamma$), shown for two travel-time assumptions (short and long). 
Table~\ref{tab:statistics_solar_assoc} (Appendix~\ref{annex: timings_distrubtion}) gives counts of positive and negative delays under both assumptions. 
Notably, $\sim$21\% (25/121; short travel) and $\sim$19\% (23/121; long travel) of SEPs occur during the impulsive flare phase (after SXR onset but before SXR peak) \citep{Benz_1993}, while $\sim$35\% (43/121) and $\sim$41\% (50/121) occur during the SXR decay, indicating substantial late or extended acceleration. 
Seven events show extreme delays ($>360$~min); these are listed in Appendix~\ref{annex: timings_distrubtion}.

\begin{figure}[t]
    \centering
    \includegraphics[width=0.99\linewidth]{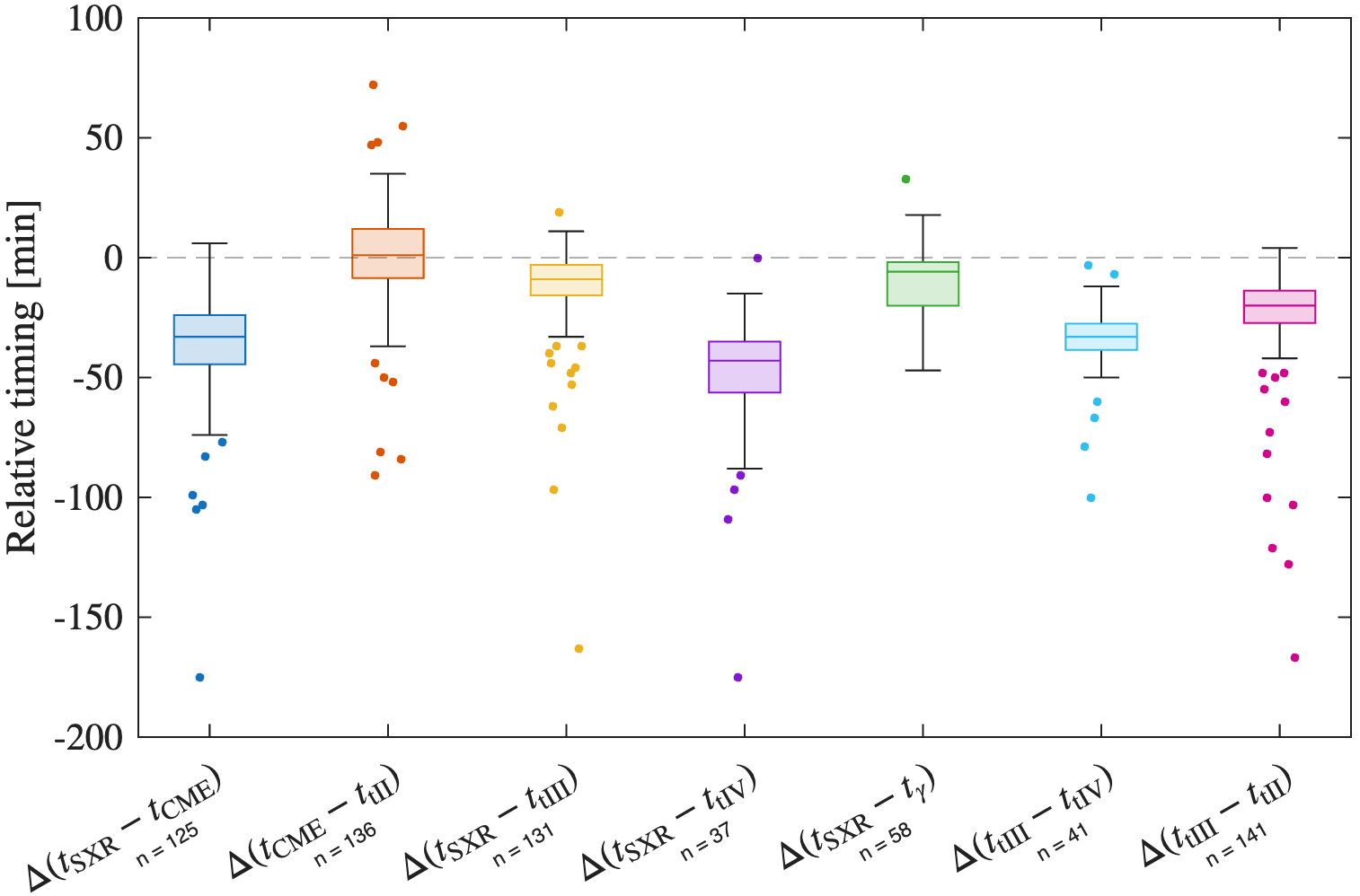}
    \caption{Same as Figure \ref{fig:sol-asso-SEP_delays} for the timing difference between solar associations in minutes.
    From left to right, comparisons between: soft X-ray onset time ($t_{\rm{SXR}}$) and CME first appearance in SOHO/LASCO coronographs ($t_{\rm{CME}}$) (blue) ; $t_{\rm{CME}}$ and type II radio burst onset time ($t_{\rm{tII}}$) (red) ; $t_{\rm{SXR}}$ and type III radio burst onset time ($t_{\rm{tIII}}$) (yellow) ; $t_{\rm{SXR}}$ and type IV radio burst onset time ($t_{\rm{tIV}}$) (purple) ; $t_{\rm{SXR}}$ and gamma ray burst onset time ($t_{\rm{\gamma}}$) (green) ; $t_{\rm{tIII}}$ and $t_{\rm{tIV}}$ (cyan) ; $t_{\rm{tIII}}$ and $t_{\rm{tII}}$ (pink).}
    \label{fig:associations_delays}
\end{figure}

\begin{figure}[t]
    \centering
    \includegraphics[width=0.9\linewidth]{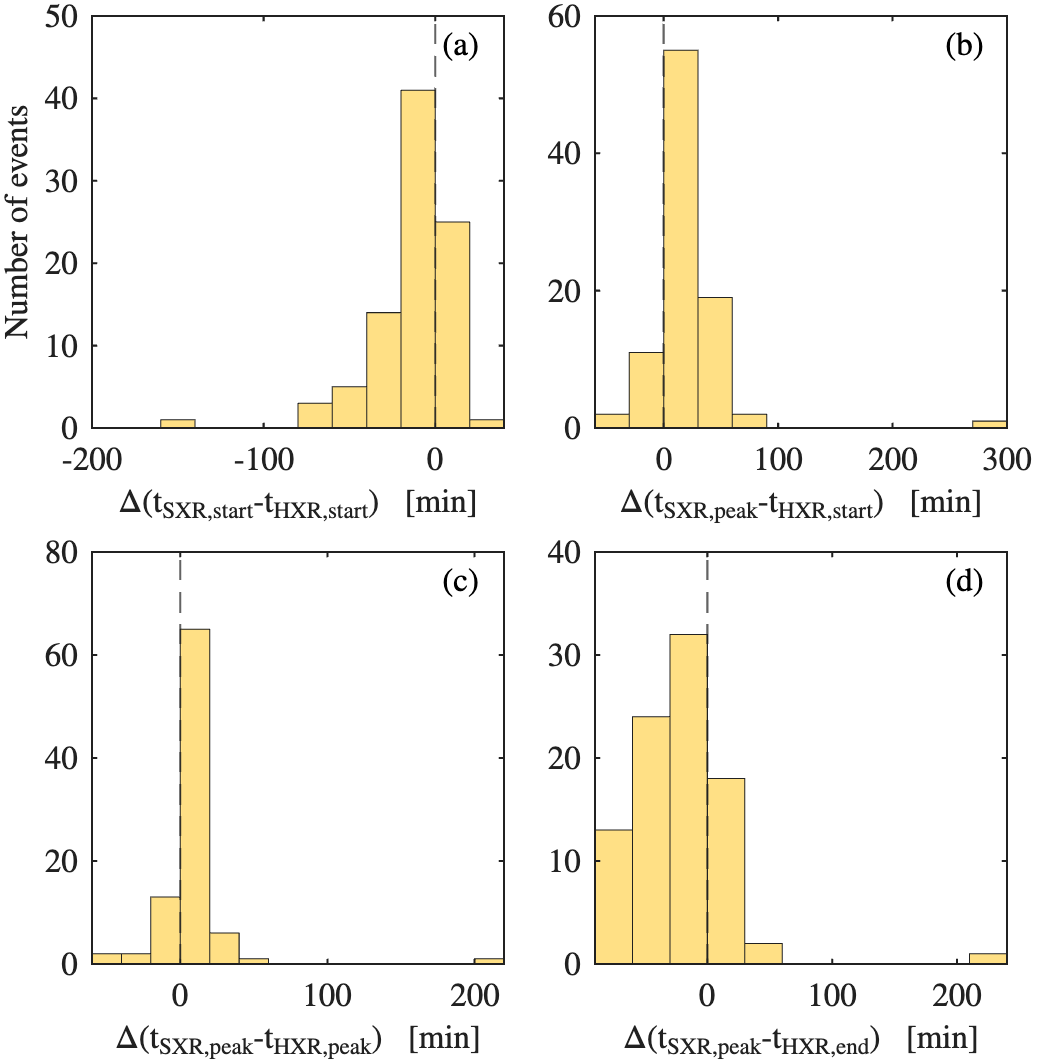}
    \caption{Timing difference (hours in UT) between soft and hard X-rays.}
    \label{fig:xr_delays}
\end{figure}

Timing between solar associations is shown in Fig.~\ref{fig:associations_delays}. 
Type~IV bursts consistently follow SXR onset and typically occur after type~III bursts. 
Coronal mass ejection first appearances and type~II onsets are tightly clustered around zero delay, consistent with CME-driven shocks as the type~II source, although individual discrepancies up to $\sim$1.5~h exist.

Two notable exceptions appear in the timing distributions.
Event~72 (17 January 2005) exhibits an unusually long SXR rising phase of about three hours, leading to a CME first appearance in LASCO which occurs few minutes ($<6$~min) before the SXR onset.
This event is also visible in the $\Delta(t_{\mathrm{SXR, onset}} - t_{\mathrm{HXR, onset}})$ distribution in Fig.~\ref{fig:xr_delays} panel (a).
There is a strong tendency for SXR onset to precede HXR one, consistent with \citet{Veronig_2002}, who found this behaviour in 92\% of events.
Event~104, on the other hand, shows a delayed SXR peak occurring $\sim$3.5~h after the CME first appearance and the other solar signatures, likely due to a second, overlapping flare episode.
It appears as an isolated point in $\Delta(t_{\mathrm{SXR}} - t_{\mathrm{tIII}})$ and $\Delta(t_{\mathrm{SXR}} - t_{\mathrm{tIV}})$ distributions (Fig. \ref{fig:associations_delays}).
Apart from these two cases, SXR and HXR peak times generally agree within 50~min, reflecting their close physical connection during the impulsive phase of energy release.

\begin{figure}[t]
    \centering
    \includegraphics[width=0.95\linewidth]{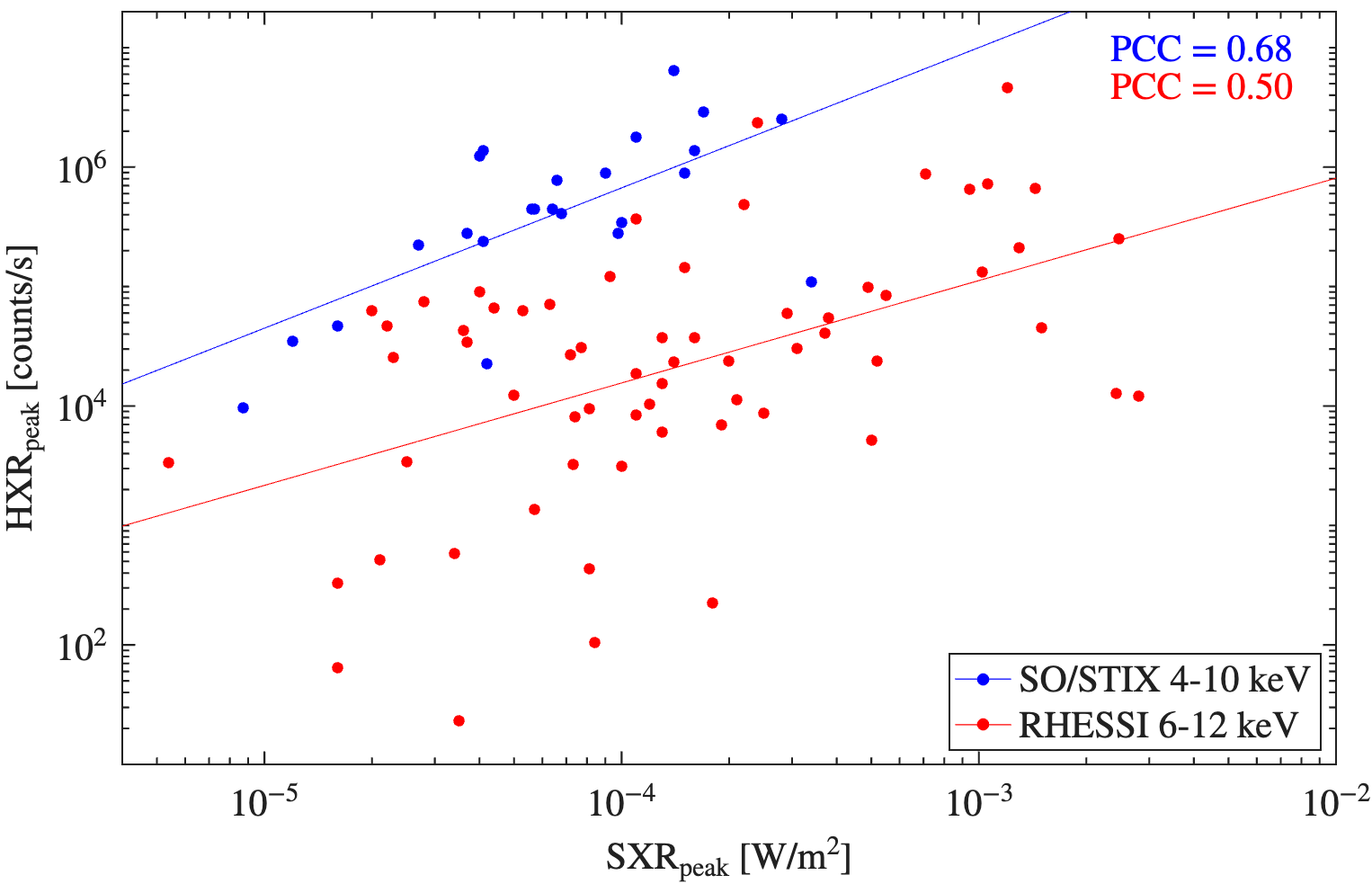}
    \caption{Hard X-ray peak intensity in counts/s from RHESSI (red points) and Solar orbiter/STIX (blue points) as a function of soft X-ray peak intensity in W/m$^2$ from GOES. The PCC of the respective best fits (red and blue lines) are indicated in the legend.}
    \label{fig:HXR_vs_SXR}
\end{figure}

Figure~\ref{fig:HXR_vs_SXR} compares HXR and SXR peak intensities using separate RHESSI and SO/STIX samples (see Sect.~\ref{subsec: cat_HXR}).
In this and all subsequent bi-logarithmic figures, the reported correlation coefficients and regression lines are obtained from log–log linear fits.
Power-law fits yield
\begin{align}
    \mathrm{HXR}_{\mathrm{RHESSI}} &= 10^{7.6 \pm 1.47}\,\mathrm{SXR}^{0.86 \pm 0.38}, \label{eq:hxr_vs_sxr_rhessi}\\
    \mathrm{HXR}_{\mathrm{SO/STIX}} &= 10^{10.52 \pm 2.38}\,\mathrm{SXR}^{1.18 \pm 0.56}, \label{eq:hxr_vs_sxr_solo}
\end{align}
with Pearson correlation coefficients (PCCs) of 0.50 (N=65, p$=2.8\times10^{-5}$, 95\% confidence interval (CI) [0.28,0.66]) for RHESSI and $0.68$ (N=25, p$=2.8\times10^{-4}$, 95\% CI [0.37,0.85]) for SO/STIX.
These results are comparable to earlier studies \citet[e.g.][, PCC = 0.57]{Veronig_2002b}.
However, correlation between SXR peak and HXR fluence (not shown here) are weaker (PCC of 0.35 (N=64, p$= 4.1\times 10^{-3}$, 95\% CI [0.11, 0.55]) and 0.46 (N=23, p$=2.6\times 10^{-2}$, 95\% CI [0.06, 0.73]), respectively), compared to the 0.71 obtain by \citet[]{Veronig_2002b}. 

\begin{figure}[t]
    \centering
    \includegraphics[width=0.95\linewidth]{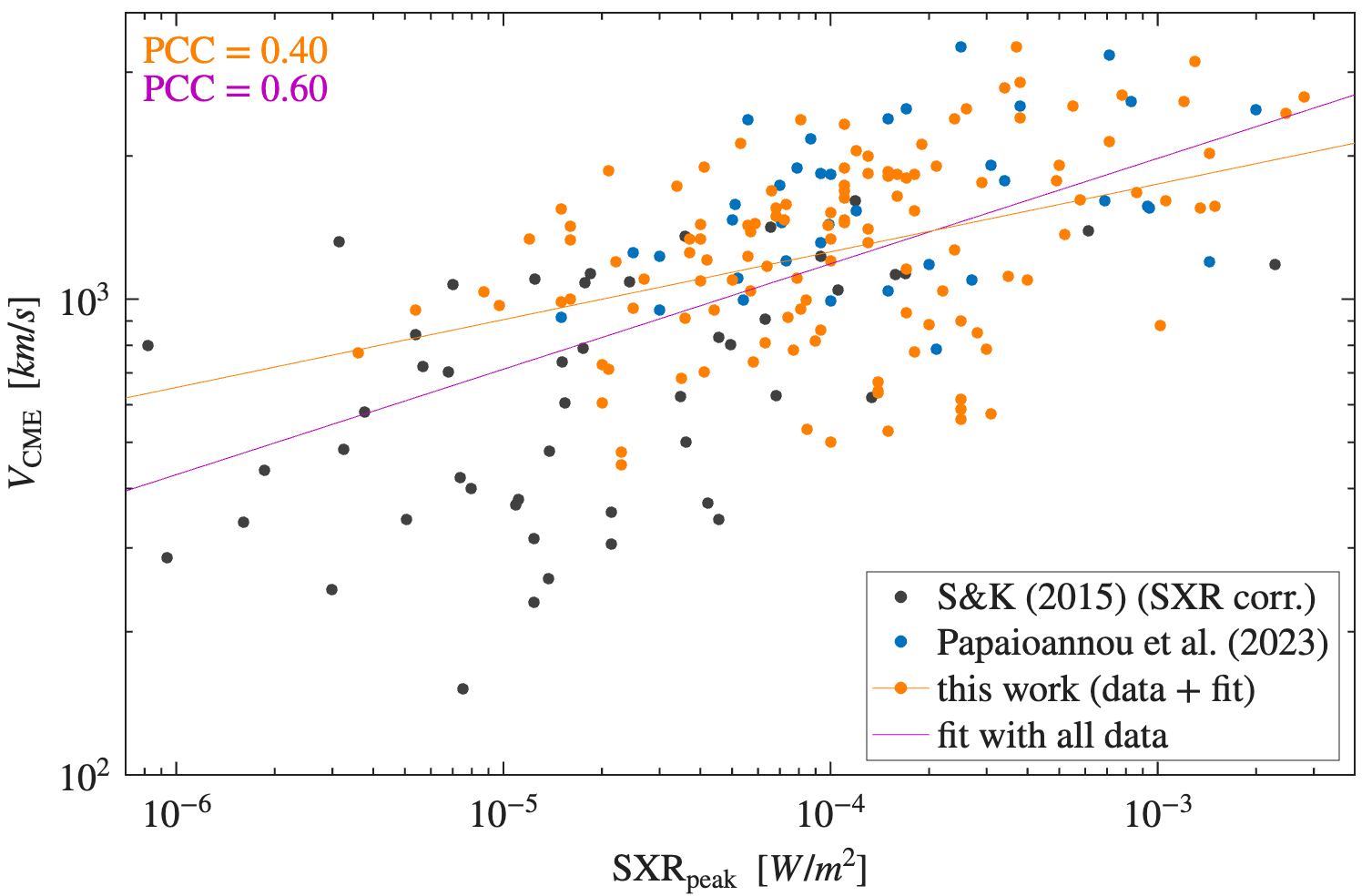}
    \caption{Coronal mass ejection speed from SOHO/LASCO as a function of the soft X-ray peak from GOES, for 49 events studied in \citet{Salas-Matamoros_2015} (grey), 65 SEP events that reach E>100 MeV studied by \citet{Papaioannou_2023} (blue), and 124 events from the present catalogue (orange). Power law fits for our sample and the combined three are lines in orange and pink, respectively.}
    \label{fig:CME_vs_SXR}
\end{figure}

Figure~\ref{fig:CME_vs_SXR} shows CME speed versus SXR peak flux. For our catalogue:
\begin{equation}
    V_{\mathrm{CME,\,this\,work}} = 10^{3.67 \pm 0.23}\,\mathrm{SXR}_{\mathrm{max}}^{0.14 \pm 0.06},
    \label{eq_cme_afo_sxr_us}
\end{equation}
with PCC = 0.40 (N=124, p$= 3.3\times 10^{-6}$, 95\% CI [0.25, 0.54]).
Combining our sample with literature datasets \citep[i.e.][]{Salas-Matamoros_2015, Papaioannou_2023} yields
\begin{equation}
    V_{\mathrm{CME}} = 10^{3.96 \pm 0.17}\,\mathrm{SXR}_{\mathrm{max}}^{0.22 \pm 0.04},
    \label{eq_cme_afo_sxr}
\end{equation}
with PCC = 0.60 (N=238, p$= 6.1\times 10^{-22}$, 95\% CI [0.50, 0.68]).
These results are in agreement with previous studies that reported PCC of 0.43 \citep{Papaioannou_2024} and 0.48 \citep{Salas-Matamoros_2015} and within the range of 0.37 and 0.50 that has been identified in E$>$10 MeV SEPs \citep{Yashiro_2009} and GLEs \citep{Gopalswamy_2012}, respectively.
The modest correlation indicates that, although flares and CMEs share a common origin in the large-scale reconfiguration of the coronal magnetic field and therefore do not occur in isolation \citep[see the relevant discussion in][]{Papaioannou_2025b}, CME kinematics are governed by additional factors. In particular, the CME speed is strongly modulated by the properties of the ambient coronal environment and by the reconnection dynamics that drive energy release. Consequently, the soft X-ray flux, while a useful proxy for flare intensity, cannot by itself provide a substitute of the CME speed. Furthermore, it is evident that the related SXR fluxes and the corresponding CME speeds across all samples at E$>$100 MeV in Fig.~\ref{fig:CME_vs_SXR} (orange and blue points) indicate that such high-energy SEP events are associated with strong flares and fast CMEs. Given our pre-selection of highly energetic events, it is normal to observe a shift of the spread relative to the \citet{Salas-Matamoros_2015} sample (black points).

\begin{figure}[t]
    \centering
    \includegraphics[width=0.95\linewidth]{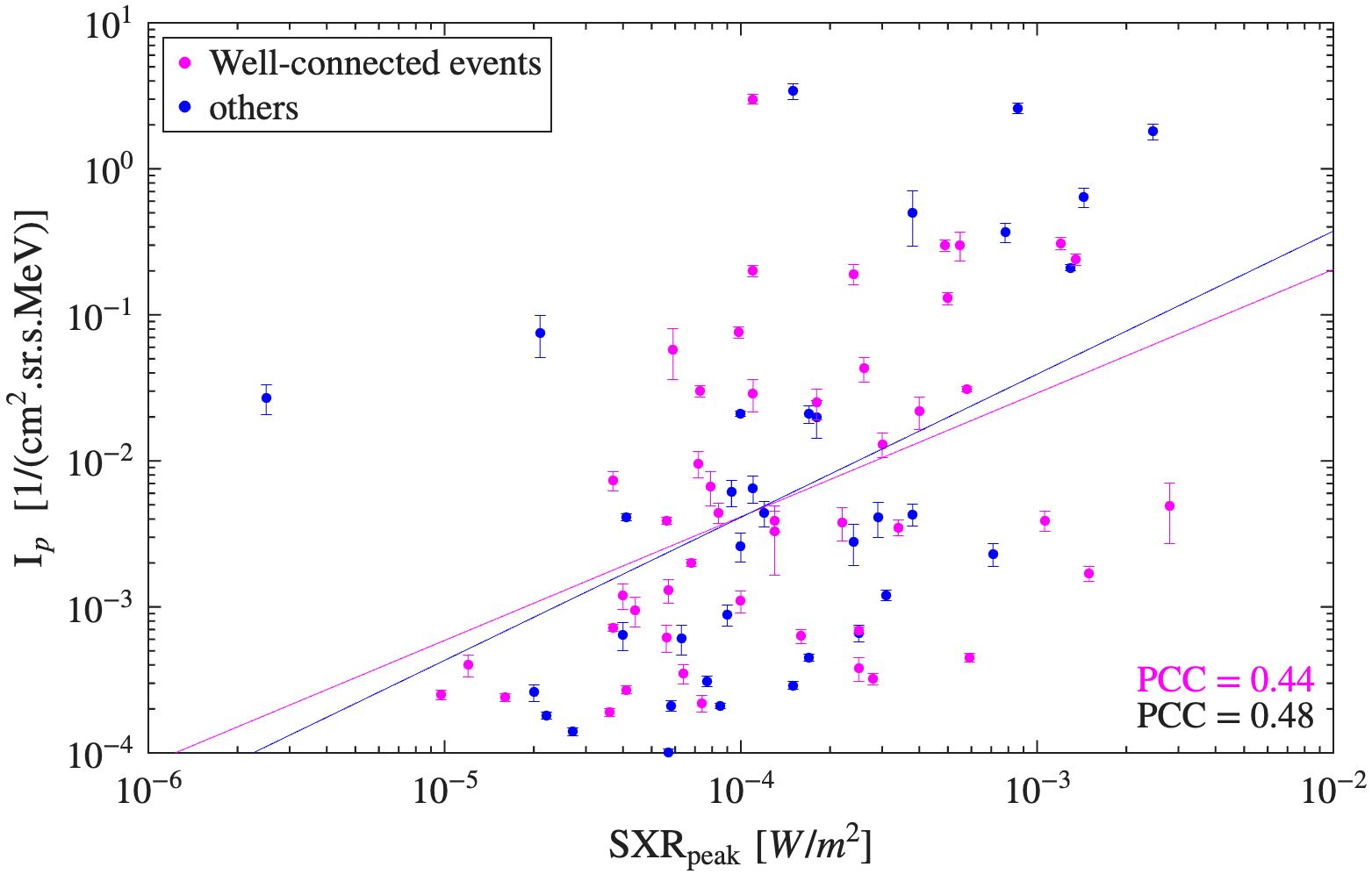}
    \caption{SOHO/EPHIN proton peak intensity $>$100~MeV (I$_p$) with uncertainties as a function of the soft X-ray peak from GOES. Events considered as well connected, i.e with a SXR longitude between 20 and 87 degrees, are shown in pink. Pearson correlation coefficients of well-connected and all events are, respectively, in pink and black.}
    \label{fig:peakflux_vs_SXR}
\end{figure}

Proton peak intensities ($>$100~MeV, EPHIN) correlate moderately with SXR peak flux (Fig.~\ref{fig:peakflux_vs_SXR}; PCC = 0.48 (N=84, p$= 4.6\times 10^{-6}$, 95\% CI [0.29, 0.63])), consistent with the 0.49 reported by \citet{Papaioannou_2016}. 
Restricting to magnetically well-connected longitudes (20°–87°, following \cite{Belov_2005}) yields PCC = 0.44 (N=49, p$= 1.8\times 10^{-3}$, 95\% CI [0.18, 0.64]), showing that improved nominal connection does not substantially strengthen the SXR–SEP relation which remains moderate.
For a given flare magnitude, the peak proton flux measured by SOHO varies by over four orders of magnitude, highlighting the large intrinsic variability of SEP events and the need for multiple parameters to explain the underlying physical processes.
This scatter suggest that SXR flux alone cannot account for the observed SEP variability; CME kinematics, shock acceleration efficiency and interplanetary transport effects likely contribute significantly.
The higher PCC (0.66) reported by \citet{Belov_2017} under similar longitudinal conditions likely reflects differences in sample definition and size.
The substantial scatter in both datasets introduces statistical noise that can weaken the apparent correlation strength.

\begin{figure}[t]
    \centering
    \includegraphics[width=0.95\linewidth]{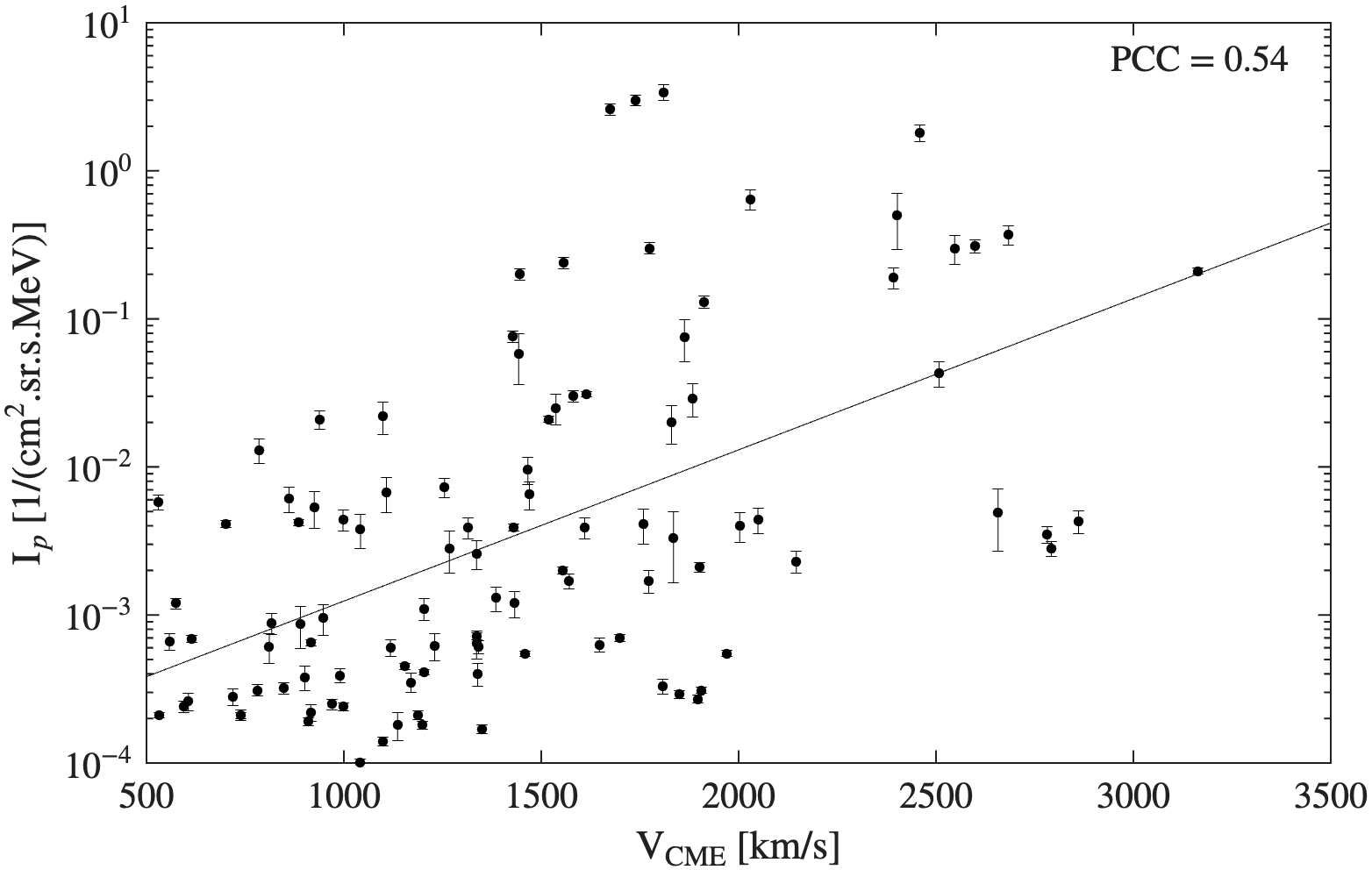}
    \caption{SOHO/EPHIN proton peak intensity $>$100~MeV (I$_p$) with uncertainties as a function of the linear CME speed measured with SOHO/LASCO (V$_{\mathrm{CME}}$).}
    \label{fig:peakflux_vs_CMEspeed}
\end{figure}

A somewhat stronger correlation is found between SEP peak flux and CME linear speed (Fig.~\ref{fig:peakflux_vs_CMEspeed}):
\begin{equation}
    \mathrm{I}_{\mathrm{p}} = 10^{-3.93 \pm 0.49}\,\mathrm{V}_{\mathrm{CME}}^{(10.21 \pm 3.12)\times10^{-4}}.
    \label{eq:sep_afo_cme}
\end{equation}
The obtained PCC of 0.54 (N=106, p$=2.8\times10^{-9}$, 95\% CI [0.38, 0.66]) is stronger than the $\sim$0.40 reported for E$>$100~MeV events \citep{Papaioannou_2016, Papaioannou_2024}, and weaker than the 0.71 found by \citet{Richardson_2014} for 25~MeV protons.
These trends align with \citet{Ameri_2024}, who showed that lower-energy SEPs (E$<$100~MeV) are primarily CME-driven, while higher-energy particles (E$>$100~MeV) likely originate from flare reconnection or low-coronal shock acceleration, possibly followed by re-acceleration at CME-driven shocks higher in the corona.
These empirical correlations must also be considered in the context of the Big Flare Syndrome \citep[BFS,][]{Kahler_1982}, whereby large eruptive events tend to produce strong signatures across many observables.
The moderate SEP-SXR correlations, and even the stronger SEP–CME relationship, may partly reflect the overall event magnitude rather than direct causal links, so these results could be interpreted as statistical associations rather than strict evidence of causation.

\begin{figure}[t]
    \centering
    \includegraphics[width=0.95\linewidth]{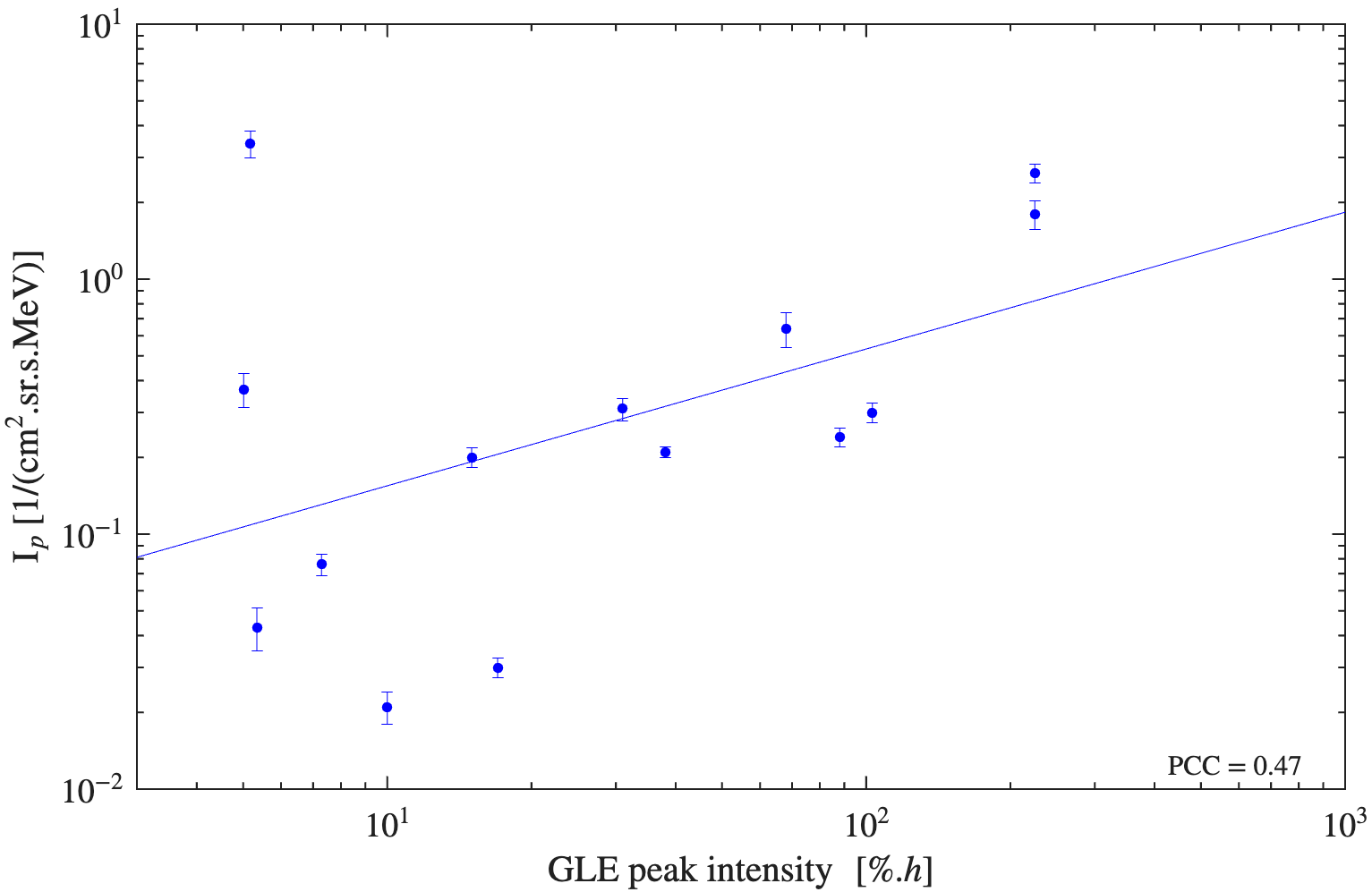}
    \caption{SOHO/EPHIN proton peak intensity $>$100~MeV (I$_p$) with uncertainties as a function of the integral GLE peak intensity (\%.h) for 14 events of the present catalogue.}
    \label{fig:glepeak_vs_seppeak}
 \end{figure}

\begin{figure}[t]
    \centering
    \includegraphics[width=0.95\linewidth]{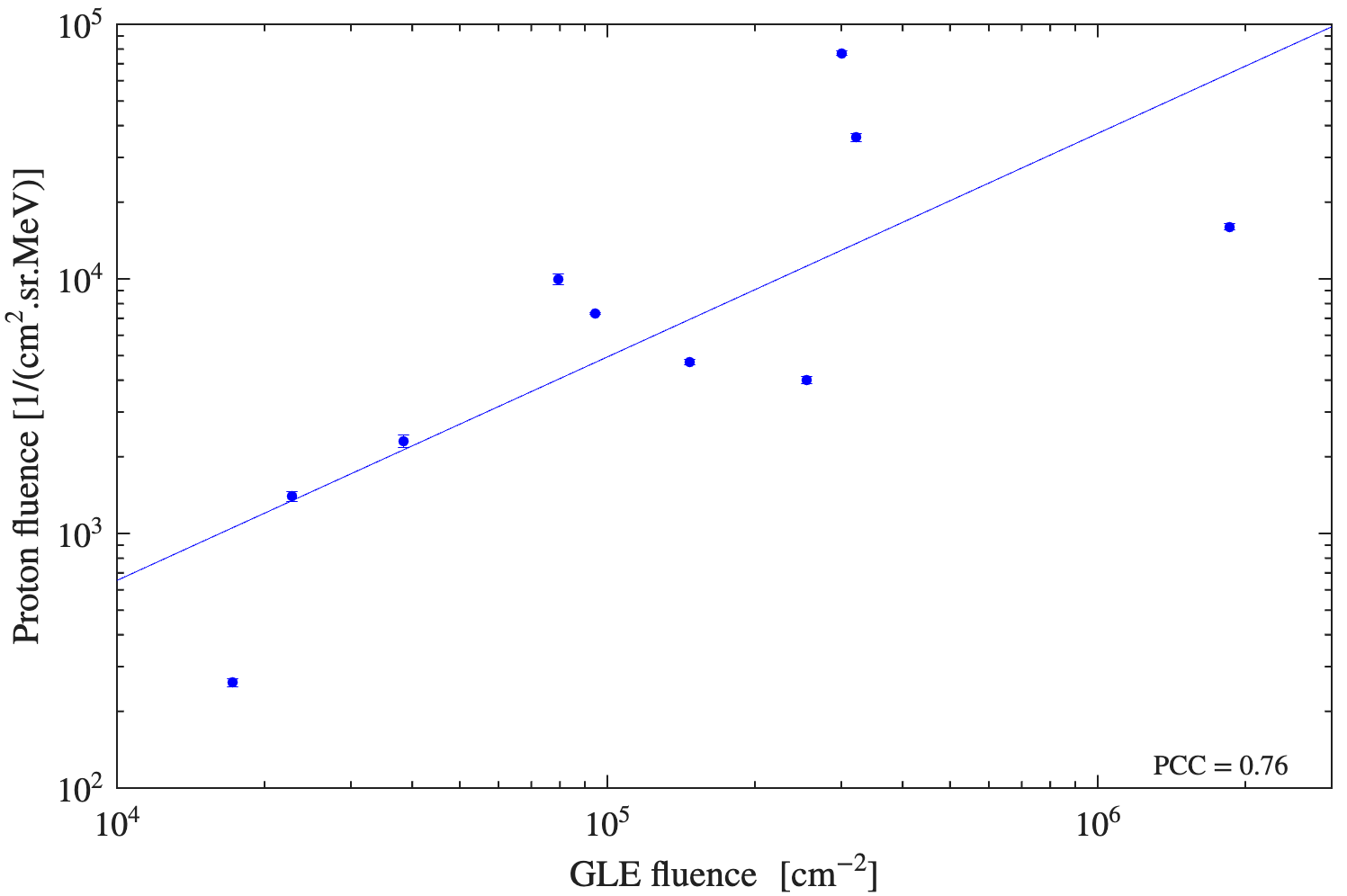}
    \caption{SOHO/EPHIN proton fluence with uncertainties as a function of the omnidirectional integral GLE fluence expressed in cm$^{-2}$ for 10 events of the present catalogue.}
    \label{fig:glefluence_vs_sepfluence}
\end{figure}

Comparisons between GLEs and SEPs peak intensities are shown in Fig.~\ref{fig:glepeak_vs_seppeak}.
For 14 GLE and sub-GLE events, we find a moderate correlation (PCC = 0.47, p=$9.2\times10^{-2}$, 95\% CI = [-0.08, 0.80]), weaker than the high-energy results of \citet{Waterfall_2023} (0.879-0.925 at $\sim$340-730~MeV) and \citet{Oh_2010} (0.867-0.874 at $\sim$380-700~MeV).
Extending the comparison to fluences (11 events; Fig.~\ref{fig:glefluence_vs_sepfluence}) yields a stronger PCC of 0.76 (p = $1.1\times10^{-2}$, 95\% CI = [0.24, 0.94]), confirming the clear relationship between high-energy SEP fluences and neutron monitor responses, consistent with earlier findings \citep{Oh_2010, Waterfall_2023}, and further reinforces the strong physical connection between space-based SEP measurements and ground-level enhancements.

\section{Conclusion}
\label{sec: conclusion}

We have compiled a comprehensive catalogue of SEP events with proton energies exceeding 100~MeV, based on a systematic search of SOHO/ERNE/HED penetrating-particle rates over May~1996–August~2024, as part of the SPEARHEAD project.
The catalogue includes 172 SEP events and is currently available in Zenodo \footnote{\url{https://zenodo.org/records/17520213}} \citep{Jarry_2025}.
Peak fluxes and fluences were derived from SOHO/EPHIN data, whose recent recalibration enables reliable energy-resolved spectra above 100~MeV.
The resulting dataset contains 172 events with complete information on their solar associations, including soft and hard X-rays, radio bursts of types~II–IV, sustained $>$100~MeV $\gamma$-ray emission, and CME signatures.
Associations were found for nearly all events: 76\% with SXR flares, 95\% with type~II bursts, 53\% with type~IV continua, 65 of 87 Fermi/LAT-era events with $\gamma$-rays, and 96\% with CMEs—typically fast and wide, often halo CMEs.

The statistical analysis of temporal and energetic relationships between SEPs and their solar drivers reveals several key results:
\begin{itemize}[label=\textbullet]
    \item The catalogue includes 172 events from 1996–2024, with near-complete coverage of CMEs, flares, and radio bursts, providing the most comprehensive reference to date for high-energy ($>$100 MeV) SEP events observed by SOHO and enabling cross-calibration with other missions.
    \item Timing analyses confirms that most $>$100 MeV SEP releases occur shortly after X-ray onsets and CME first appearances (Fig.~\ref{fig:sol-asso-SEP_delays}), indicating nearly simultaneous release of flare- and shock-accelerated particles.  
    A few outliers (Appendix~\ref{annex: timings_distrubtion}) reveal complex multi-phase or overlapping eruptions, illustrating the diversity of acceleration scenarios.
    \item Distributions of solar timing differences (Figs.~\ref{fig:associations_delays}, \ref{fig:xr_delays}) show consistent sequencing: hard X-rays and type III bursts typically follow SXR onsets, confirming that impulsive acceleration is closely tied to early reconnection signatures.
    \item The comparison of soft and hard X-ray timings and peaks (Figs.~\ref{fig:xr_delays}, \ref{fig:HXR_vs_SXR}) supports the Neupert effect, with SXR emission rising gradually after HXR impulsive phases.  
    Correlations between their peak intensities (PCC = 0.50-0.68) confirm the expected impulsive–thermal coupling in major SEP-producing flares \citep{Veronig_2005}.
    \item The first appearance of a CME usually follows the SXR onset by 30–60 min, consistent with a shared magnetic restructuring.  
    Correlations between CME speed and SXR peak flux remain moderate (Fig.~\ref{fig:CME_vs_SXR}, PCC = 0.4-0.6), in line with previous studies \citep{Yashiro_2009, Salas-Matamoros_2015, Papaioannou_2024}, consistent with the idea that CME kinematics are influenced by both the flare energy release and other factors such as ambient coronal conditions and reconnection geometry.
    \item Correlation between proton intensity and SXR peak flux remain moderate (Fig.~\ref{fig:peakflux_vs_SXR}), and the large scatter implies that SEP intensity is controlled by multiple factors beyond simple flare magnitude.
    The correlation with CME speed (Fig.~\ref{fig:peakflux_vs_CMEspeed}, PCC = 0.54) is stronger than reported in previous high-energy studies \citep{Papaioannou_2016}, but weaker than those found for lower-energy SEPs \citep{Richardson_2014}.
    These empirical associations are consistent with scenarios in which both flare processes and CME-driven shocks contribute to $>$100 MeV SEPs, although they may partly reflect overall event size (BFS) rather than a direct link.
    In addition, magnetic connectivity and pre-conditioning are important modifiers of high-energy SEP-event characteristics.
    \item Comparisons between $>$100 MeV SEPs and GLEs (Figs.~\ref{fig:glepeak_vs_seppeak}–\ref{fig:glefluence_vs_sepfluence}) reveal a clear link between space- and ground-based measurements, with moderate correlation for peak fluxes (PCC = 0.47) and stronger for fluences (0.76), consistent with \citet{Oh_2010, Waterfall_2023}.  
    This confirms that the most energetic SEPs observed in space are the direct counterparts of GLEs.
\end{itemize}

Future developments of the catalogue are underway. The inclusion of CME observations from the STEREO/SECCHI suite \citep{Howard_2008}—particularly COR1, COR2, and HI imagery—will be consider as well as additional CME properties, such as de-projected speeds from the HELCATS catalogue\footnote{\url{https://www.helcats-fp7.eu/catalogues/wp3_cat.html}}, to improve association accuracy and analysis. Complementary SXR observations from the MESSENGER X-Ray Spectrometer \citep{Schlemm_2007}, when the spacecraft viewed the solar far side, and improved magnetic connectivity estimates using IRAP’s Magnetic Connectivity Tool \citep{Rouillard_2020}, will further refine the association of SEPs with their solar sources.

High-energy SEPs ($>$100 MeV) probe the most energetic regimes of particle acceleration associated with flares and CME-driven shocks.
Associated with the more abundant lower-energy SEPs (E$>$10 MeV), which provide essential and complementary constraints on acceleration and transport, sampling different parts of the particle spectrum and responding differently to shock strength, turbulence, and seed populations, these highest-energy particles offer additional diagnostic power.
The rapid arrival of the earliest, prompt high-energy particles which have experienced less scattering preserve key information on acceleration timing and magnetic connectivity.
Their close association with GLEs thus linked space-based SEP measurements to ground based observations.
Given their rarity and strong association with the most powerful solar eruptions, high-energy SEP events serve as sensitive indicators of extreme solar activity.
Combined with low-energy SEPs, their properties provide a more complete view of the physical conditions governing the most powerful solar events.
By consolidating nearly three decades of observations into a single, uniformly processed catalogue, this work provides the most extensive reference dataset to date for investigating high-energy SEP origins, acceleration and propagation mechanisms, and to address many of the outstanding questions in SEP research.

\section*{Data availability}

The complete catalogue created and used in
this study is provided on \url{https://zenodo.org/records/17520213}. An interactive version of this catalogue is available at \url{https://spearhead-he.eu/catalogues_hub?prefilter=soho100mev}.

\begin{acknowledgements}
    We thank the anonymous referee for their thorough and accurate review, which greatly improved the manuscript.
    This research received funding from the European Union’s Horizon Europe programme under grant agreement No 101135044 (SPEARHEAD) [\url{https://spearhead-he.eu/}]. Views and opinions expressed are, however, those of the author(s) only and do not necessarily reflect those of the European Union or the European Health and Digital Executive Agency (HaDEA). Neither the European Union nor the granting authority can be held responsible for them. B.H. and M.K. further acknowledge the support of the EPHIN and SEPT from the German Federal Ministry for Economic Affairs and Energy, the German Space Agency (Deutsches Zentrum für Luft- und Raumfahrt e.V., DLR) under grants 50OC2102, 50OC3202.
\end{acknowledgements}

\bibliographystyle{aa}
\bibliography{aa57965-25}

\appendix

\section{SOHO/ERNE Penetrating Particle Counter}
\label{annex: soho-erne_PCC}

ERNE consists of two multi-layer particle telescopes, LED and HED (Low and High Energy Detector, respectively). 
The data used in this work are taken from HED. 
HED comprises a hodoscope with four layers of silicon strip detectors (each 0.3~mm thick), followed by a fifth silicon detector layer (1~mm thick, referred to as detector D1) and two scintillator layers made of CsI(Tl) (8~mm, D2) and BGO (15~mm, D3), respectively.
Below the stack lies an anti-coincidence plastic scintillator (6~mm, AC1) designed to trigger on particles penetrating the entire stack.
In particular, we use the penetrating-particle counter of HED, which primarily responds to protons with energies above 100~MeV.
The response of this channel is still being investigated using GEANT4 simulations \citep{Agostinelli_2003}, as well as through cross-calibration with other datasets.
For this reason, the first release of the catalogue relies on counting rates rather than on particle intensities derived from these rates.

\begin{figure}[h!]
    \centering
    \includegraphics[width=\linewidth]{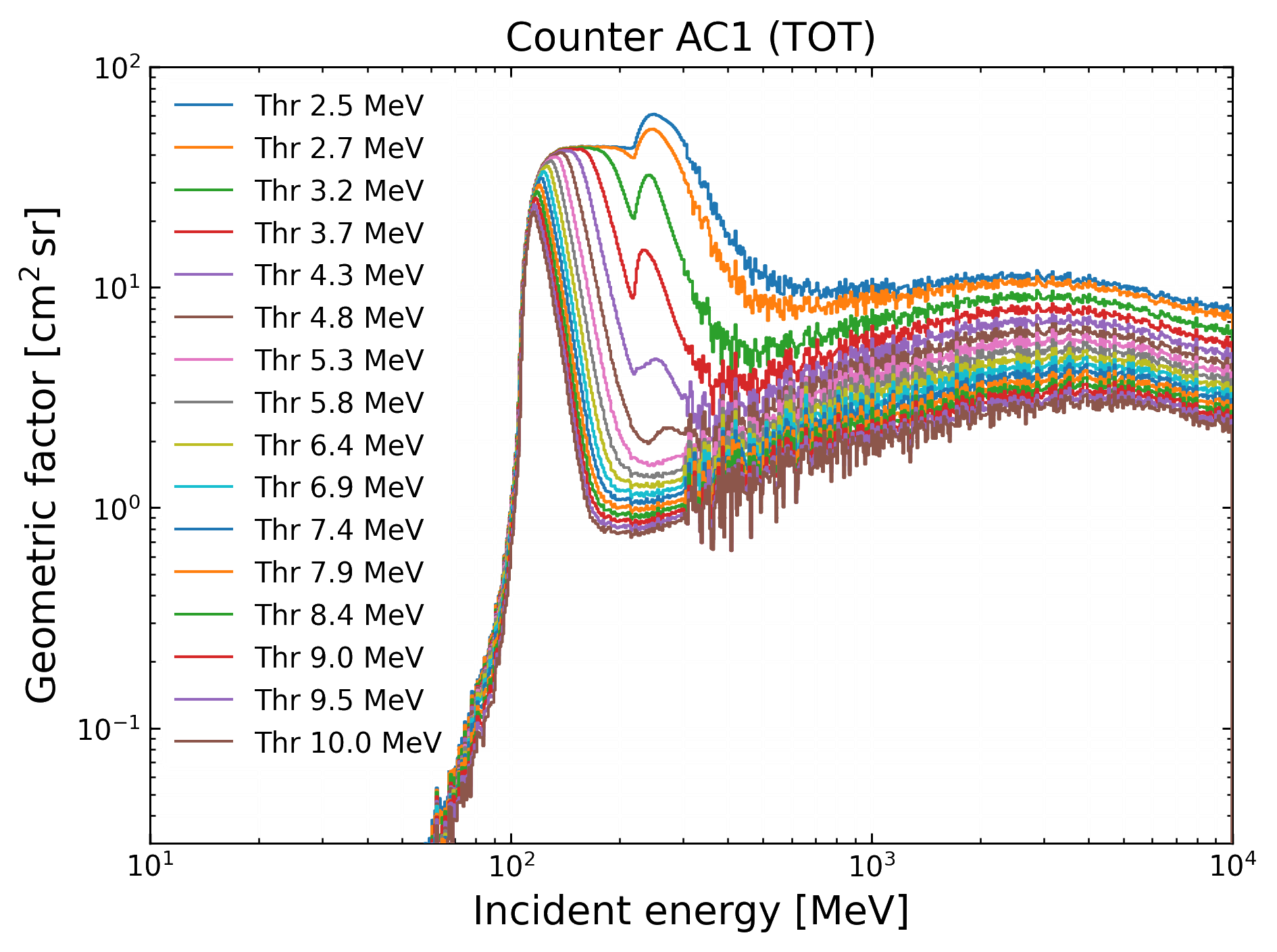}
    \caption{Simulated response function of the ERNE/HED penetrating-particle counter, accounting for both forward- and backward-penetrating protons. For the latter, the spacecraft structure below the instrument is assumed to correspond to an 10~cm thick aluminium layer. The different curves represent various (unknown) trigger levels of the bottom plastic scintillator in the stack.}
    \label{fig:HED-AC1}
\end{figure}

\begin{figure}
    \centering
    \includegraphics[width=\linewidth]{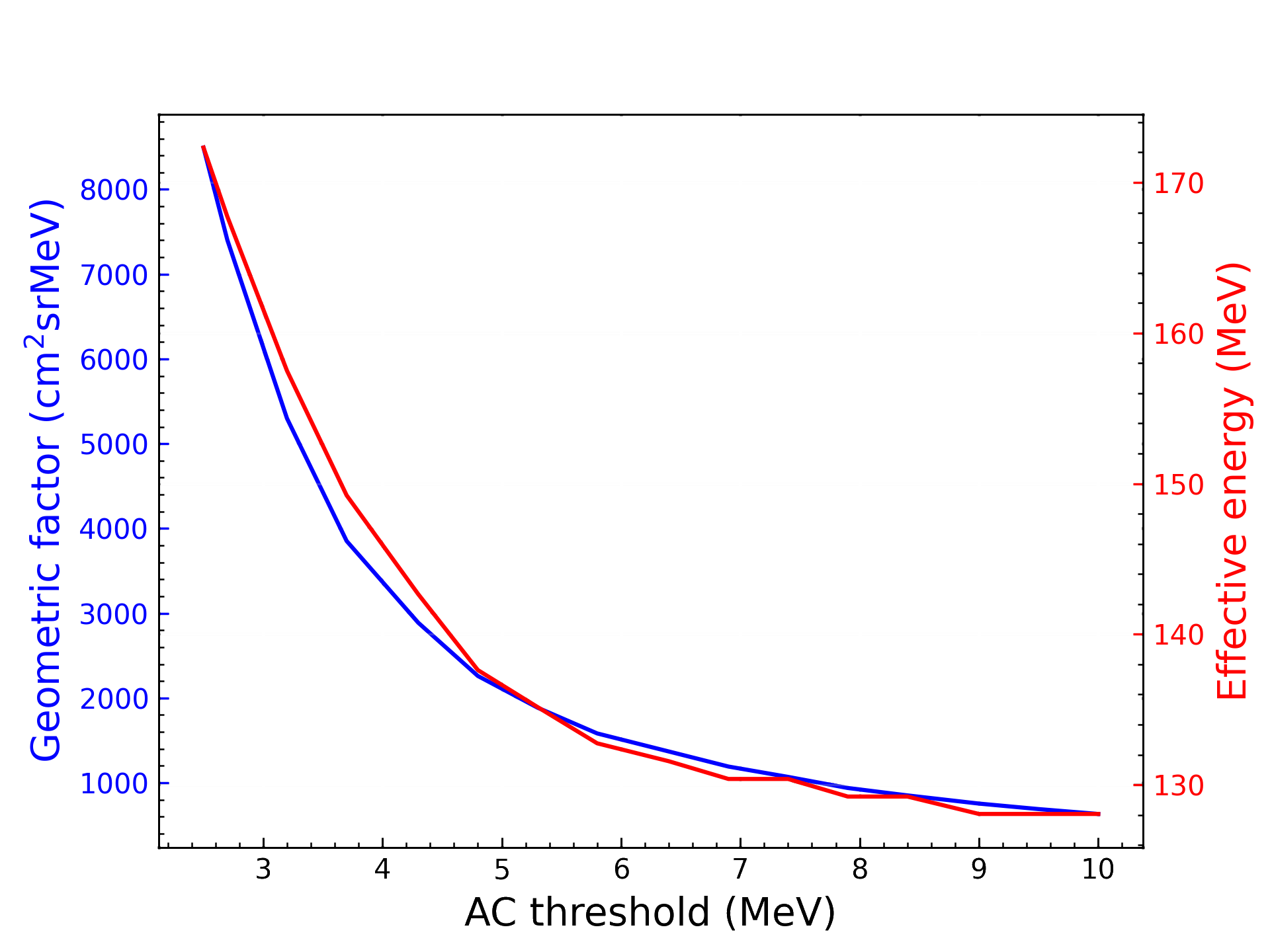}
    \caption{Geometric factor (blue) and effective energy (red) of the penetrating-particle counter as a function of the plastic scintillator detection threshold (‘AC threshold'), obtained from bow-tie analysis.}
    \label{fig:bowtie-ERNE}
\end{figure}

Preliminary response functions for protons simulated with GEANT4 are shown in Fig.~\ref{fig:HED-AC1} for different (still unknown) triggering levels of the plastic scintillator at the bottom of the stack.
It is evident that the effective energy of the counter exceeds 100~MeV, and a standard bow-tie analysis \citep[e.g.][Sect. 5.3.3]{Oleynik_2021} gives a range of 128--172~MeV for threshold levels between 2.5 and 10~MeV (Fig.~\ref{fig:bowtie-ERNE}).

\begin{figure}
    \centering
    \includegraphics[width=\linewidth]{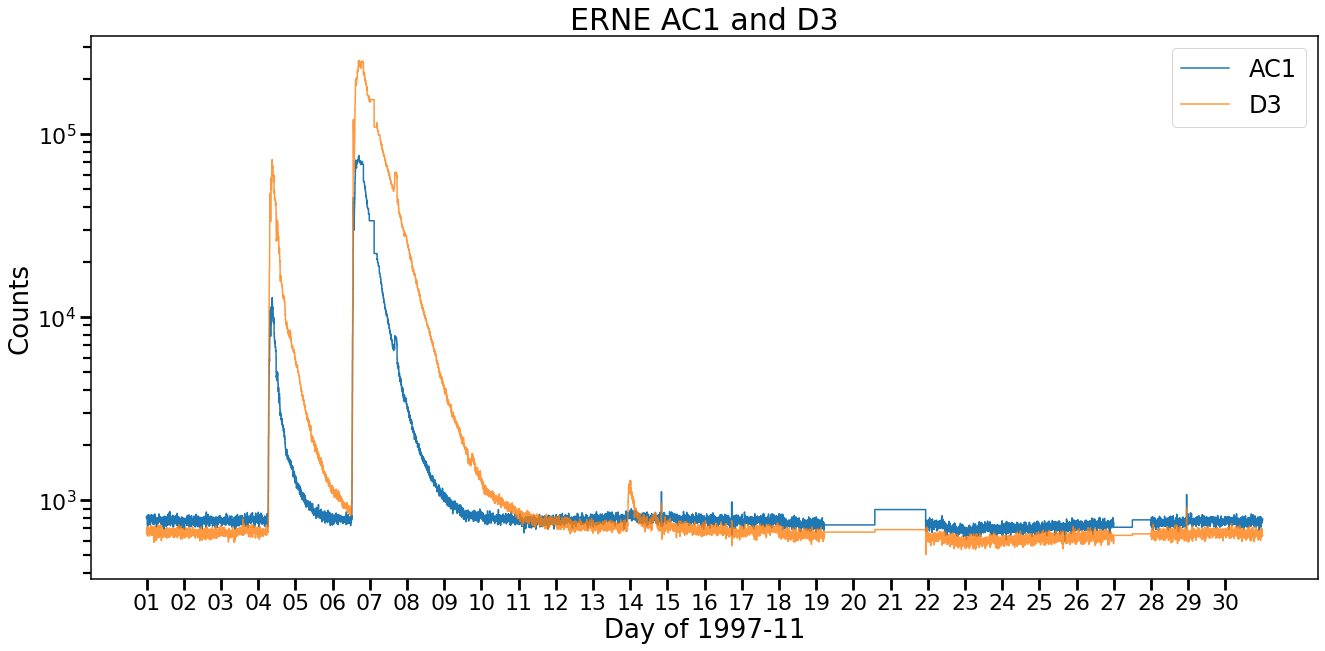}
    \caption{Example of ERNE counter observations (in counts per 5 minutes) from November 1997, showing two consecutive $>$100~MeV proton events starting on 4 and 6 November, respectively. The AC1 counter corresponds to the penetrating-particle channel, while the D3 counter measures particles stopping near the bottom of the detector stack, i.e. those not triggering the plastic scintillator.}
    \label{fig:Nov-1997}
\end{figure}

\begin{figure}[h!]
    \centering
    \includegraphics[width=1.0\linewidth]{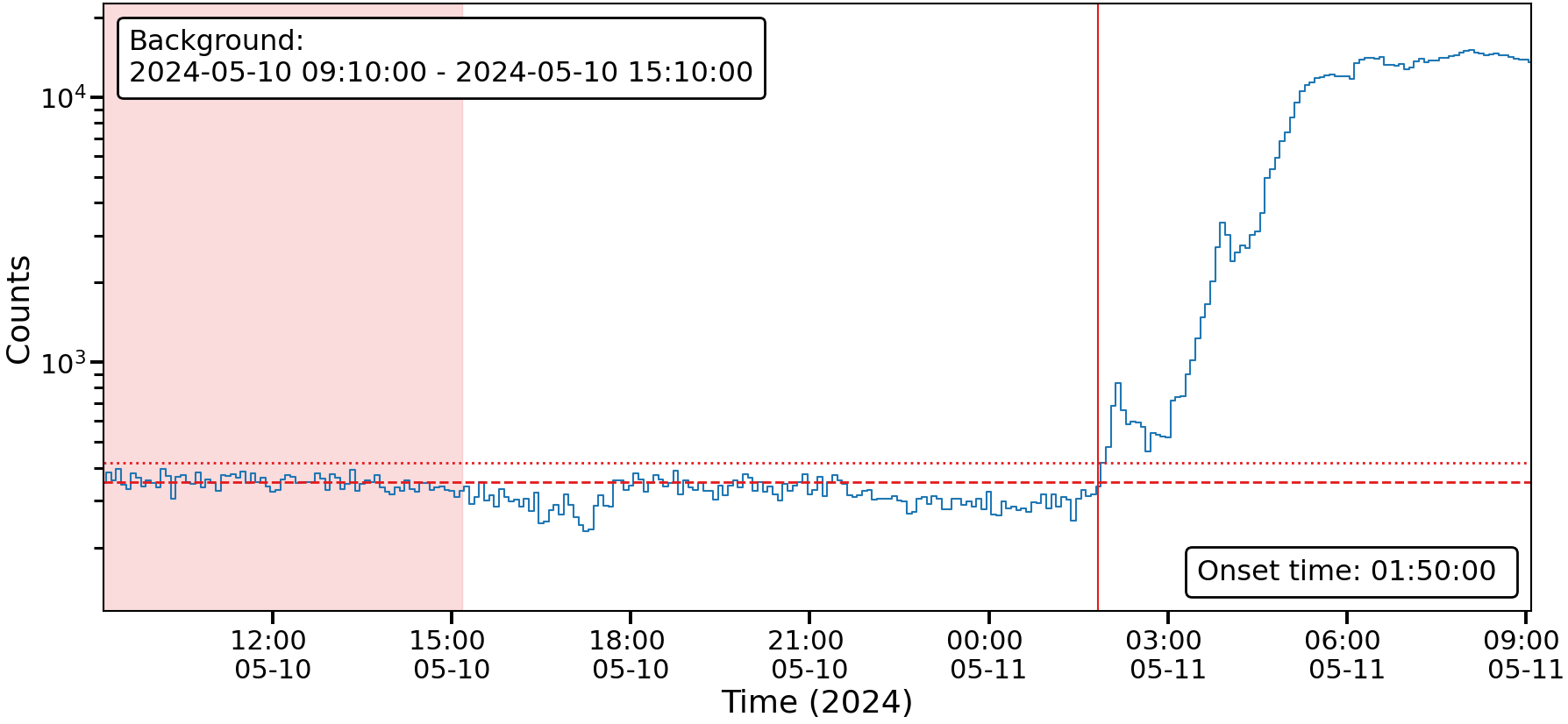}
    \caption{Detection of an event in the AC1 channel of SOHO/ERNE on 11 May 2024~01:50~UT. The pre-event background (red area) corresponds to the first six hours of the detection window. The dashed and dotted red lines indicate the background mean and the mean plus $n$ times the standard deviation ($n = 5$), respectively. The vertical line marks the detected onset time.} 
    \label{fig:ac1_event}
\end{figure}

To better distinguish SEP events from other variations in the data (see Sect.~\ref{subsec: event_detection}), we also use a counter measuring particles that stop in the D3 detector (just above AC1), which is sensitive to protons in the $\sim$50--100~MeV range.
An example time series of D3 and AC1 counters is shown in Fig.~\ref{fig:Nov-1997}.
The algorithm scans the dataset using a 24-hour moving window advanced in three-hour steps.
The first six hours of each window define the pre-event background, while the remaining 18 hours are searched for elevated count rates.
If no enhancement is detected, the window is shifted forwards by three hours; if an event is found, it is advanced by nine hours to prevent double detections.
Figure~\ref{fig:ac1_event} illustrates one example of detection (Event~No.~166, 11 May 2024~01:50~UT).
The red shading indicates the background interval, while the dashed and dotted lines show the mean background level $\mu$ and the upper level $\mu_{d} = \mu + n \cdot \sigma$ (with $n = 5$), respectively.
Note that $\mu_{d}$ is not the detection threshold itself but limits the cumulative growth of the CUSUM function \citep[see][]{Huttunen-heikinmaa_2005}.
The data gaps in ERNE observations lasting seven days or longer are as follows:

\begin{itemize}[label=\textbullet]
    \item 25 February -- 4 March 1997
    \item 24 June -- 9 October 1998
    \item 14 -- 21 November 1998
    \item 21 December 1998 -- 8 February 1999
    \item 10 -- 17 August 2001
    \item 5 -- 12 February 2002
    \item 28 February -- 11 March 2003
    \item 22 -- 29 April 2004
    \item 26 November -- 11 December 2006
    \item 29 July -- 14 August 2009
    \item 7 -- 24 September 2009
    \item 30 November 2011 -- 5 January 2012
    \item 28 January -- 10 February 2012
    \item 9 December 2012 -- 31 January 2013
    \item 29 October -- 4 December 2013.
\end{itemize}

Some of the gaps listed by \citet{Paassilta_2017} may contain short data intervals but are still treated as gaps for practical reasons, i.e. because meaningful event detection within them would not be possible.
Since 2004, shorter and more frequent interruptions have also occurred, mainly due to the reduced telemetry rate of the SOHO spacecraft during so-called ‘keyhole periods' (see Sect.~\ref{subsec: sep_peak_fluence_determination}), when the pointing of the high-gain antenna is suboptimal for communications.

\section{SOHO/EPHIN high-energy protons data}
\label{annex: soho-ephin_instrum}

EPHIN measures energetic particles — protons, electrons, and helium — across a wide range of energies, from a few mega-electronvolts up to several hundred mega-electronvolts \citep[for details, see][]{Mueller-Mellin_1995}.
The energy ranges for EPHIN are given in \citet{Kuhl_2020}.
As discussed in \citet{Kuehl-etal-2015, Kuehl-etal-2016, Kuehl-etal-2017}, the $\tfrac{dE}{dx}$–$\tfrac{dE}{dx}$ method can be applied to EPHIN’s integral channel to derive proton energy spectra from about 50~MeV to beyond 1~GeV.
An extension of the EPHIN energy range for protons to above 1~GeV, based on energy-loss data from SSD A to D, has been implemented.
Using extensive GEANT4 simulations of the instrument \citep{Agostinelli_2003}, it has been shown that the initial energy of a proton penetrating EPHIN (i.e. triggering detectors A to F) can be inferred from the energy losses in SSD C and D.
However, separating forward- and backward-penetrating protons is not possible, so the contribution from backward-penetrating particles must be accounted for.
Due to the ageing of one of the thick SSDs, the method had to be adapted to use a thin SSD, namely SSD B.

\subsection{Modification of the method by \citet{Kuehl-etal-2015}}
With SSD D becoming noisy in 2017, the method described by \citet{Kuehl-etal-2015} could no longer be directly applied.
As explained by \citet{Kuehl-etal-2016}, the energy resolution of the method depends on the energy-loss resolution of the SSDs involved.
Thus, when using a total of 3300~$\mu$m instead of 8000~$\mu$m of silicon, a significant reduction in energy resolution is expected.

\begin{figure}[h!]
    \centering
    \includegraphics[width=\linewidth]{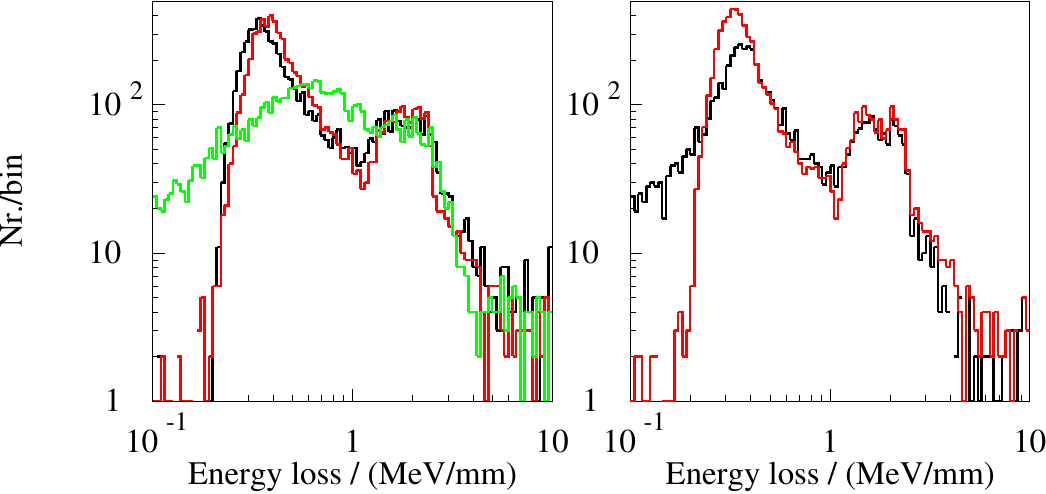}
    \caption{$\tfrac{dE}{dx}$ distributions for SSDs B (black histogram), C (red), and D (green) on the 11 May 11 2024 (GLE74). While the $\tfrac{dE}{dx}$ distributions for SSDs B and C agree well, that of SSD D differs significantly below 1.5~MeV/mm. The influence on the minima of the energy loss in SSDs B and C differs from that obtained when using SSDs D and C for energy losses below 0.5~MeV/mm.}
    \label{fig:EPHIN-Energy-Resolution}
\end{figure}

\begin{figure}
    \centering
    \includegraphics[width=0.49\columnwidth]{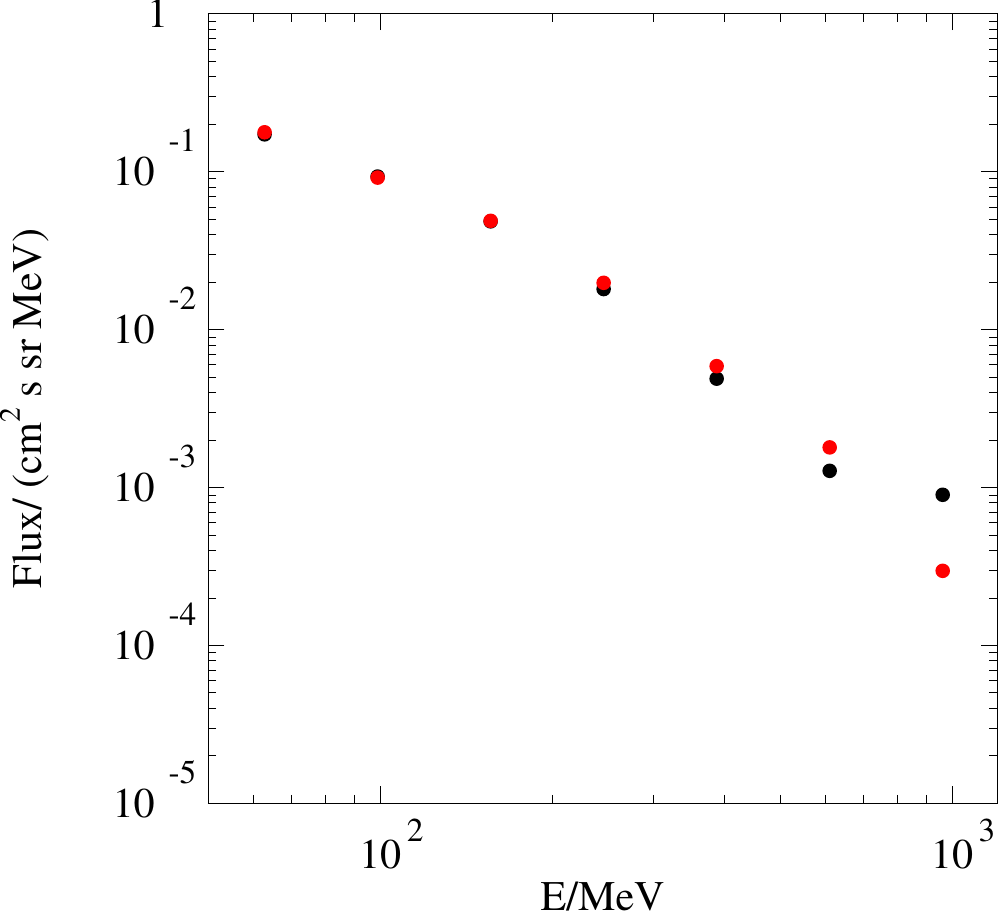}
    \includegraphics[width=0.49\columnwidth]{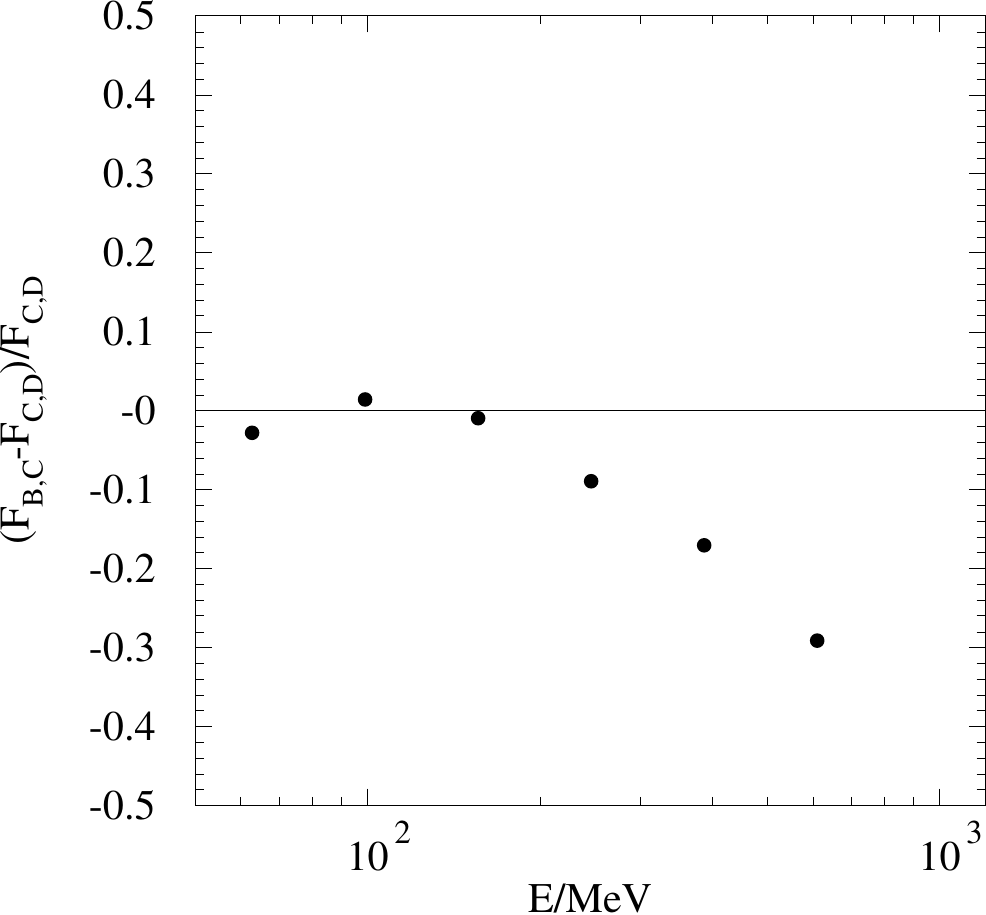}
    \includegraphics[width=0.49\columnwidth]{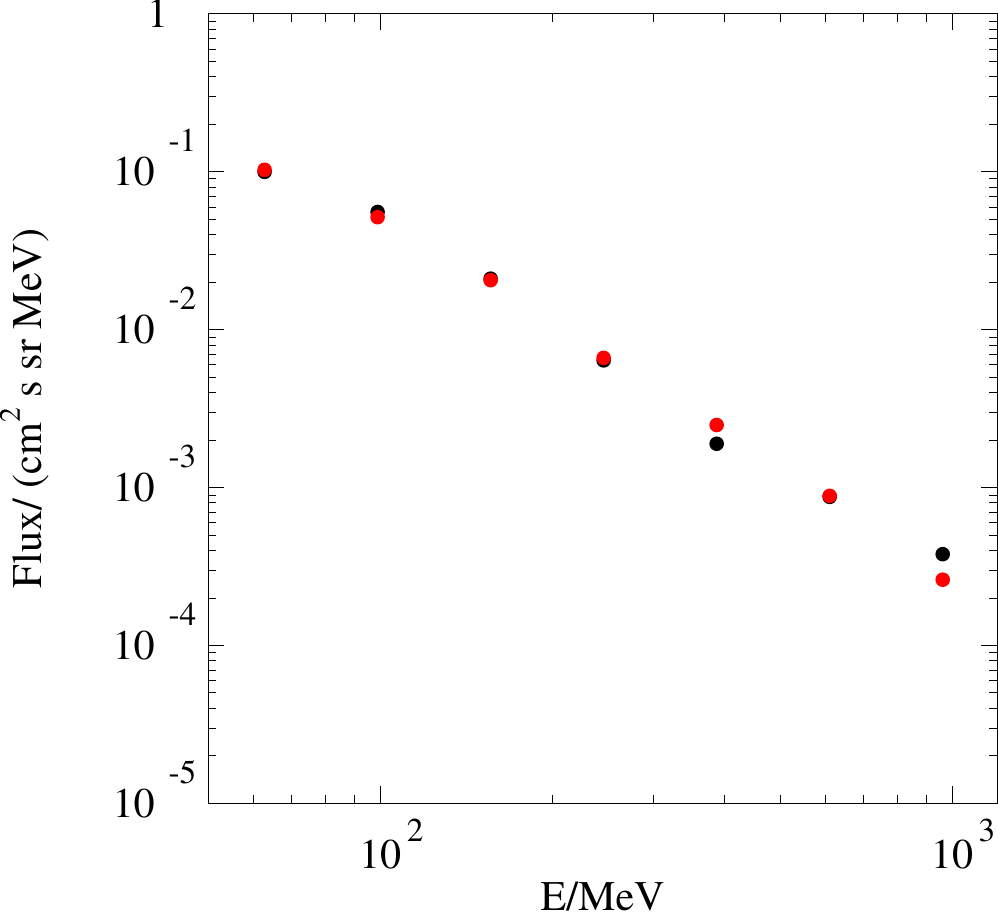}
    \includegraphics[width=0.49\columnwidth]{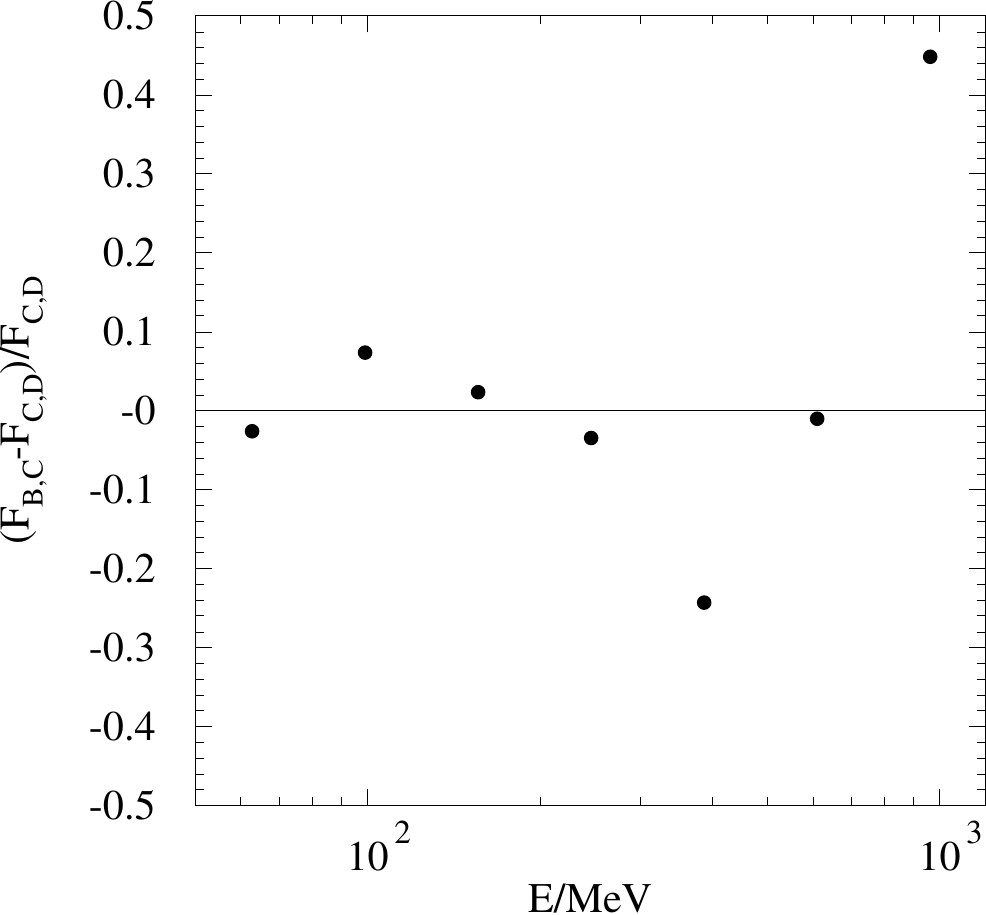}   
    \caption{Proton energy-flux spectra from \citet{Kuehl-etal-2015} (red symbols) and from this study (black symbols), and their corresponding differences (right panels). The upper and lower panels show the results using ten energy bins for GLE72 and for the sub-GLE event of 6 January 2014, respectively.}
    \label{fig:GLE-72-max}
\end{figure}

Figure~\ref{fig:EPHIN-Energy-Resolution} shows the energy-loss-per-millimetre distributions of SSDs B (black), C (red), and D (green) during the 11 May 11 2024 (GLE74).
The distribution for SSD D clearly deviates from the others, especially for $\tfrac{dE}{dx} < 1.5$~MeV/mm, where noise in SSD D strongly affects the signal.
The right panel displays the minimum energy loss per millimetre obtained from SSDs C and D (black) and from SSDs C and B (red).
Significant differences are found for $\tfrac{dE}{dx} \leq 0.5$~MeV/mm.
Figure~\ref{fig:GLE-72-max} compares the results of the method for two events reported by \citet{Kuehl-etal-2015}.
The measurements agree well within the 50–200~MeV energy range and differ by less than 30\% below 500~MeV, which is within the uncertainty estimated by \citet{Kuehl-etal-2017}.
Another observation is the reduced energy resolution, as evident when compared to the spectra presented by \citet{Kuehl-etal-2015}.

\subsection{Computation of the efficient energy and response for the $>$100~MeV proton proxy}
\begin{figure}
    \centering
    \includegraphics[width=0.5\linewidth]{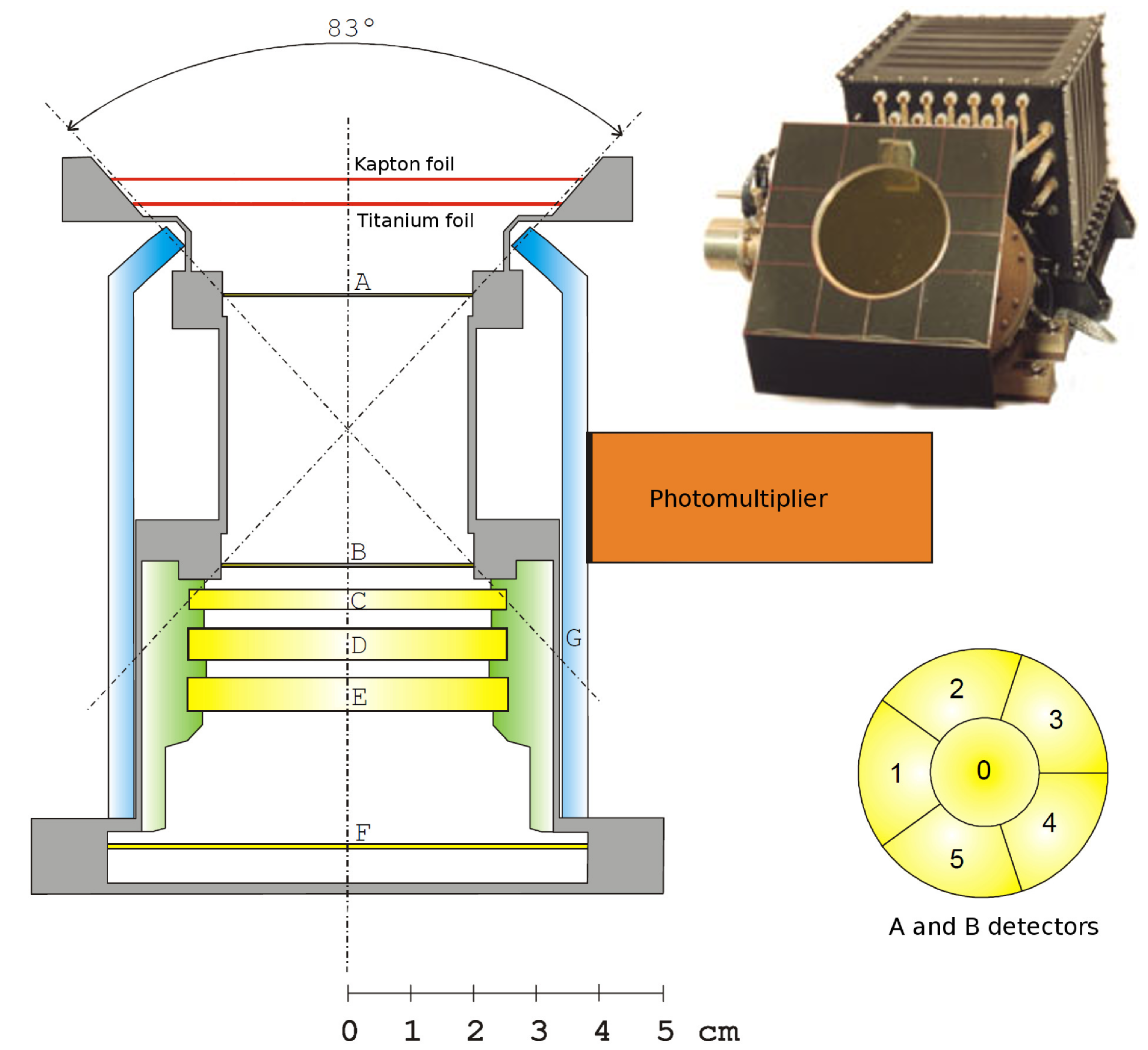}
    \caption{On the left and right a sketch and a photo of the EPHIN are shown, respectively.}
    \label{fig:EPHIN}
\end{figure}

To estimate the $>$100~MeV proton flux proxy, the energy range for protons from about 100 to 500 MeV needs to be covered to be comparable with the SOHO/ERNE channel  (see  Fig.~\ref{fig:HED-AC1}). As explained above, EPHIN utilises the $dE/dx-dE/dx$-method and relies on a precise assignment of the minimum of the two energy losses per path length in the two SSDs B and C (see Fig.~\ref{fig:EPHIN}) to the incoming energy. Due to the stochastic processes of energy loss in matter, not a single incoming energy but a distribution of different particles may contribute to a single measured energy loss.  We utilised the GEANT4 toolkit to compute the energy loss distributions for electrons in the energy range from 5 to 100~MeV and for protons and helium in the energy range from 50~MeV/amu to 2~GeV/amu (two times the rest mass of the particle), respectively. The runs were performed for particles with an energy distribution $n(E)=n_0\cdot E^{\gamma}$ and $\gamma=-1$ penetrating the instrument with an isotropic directional distribution. Since the determination of the response function is based on logarithmically spaced energy bins, $\gamma=-1$ leads to equal statistical uncertainties for each bin. Note that the influence of the power law index for SEP events is taken into account within the bow-tie method (see below).

\begin{figure}
    \centering
    \includegraphics[width=0.95\linewidth]{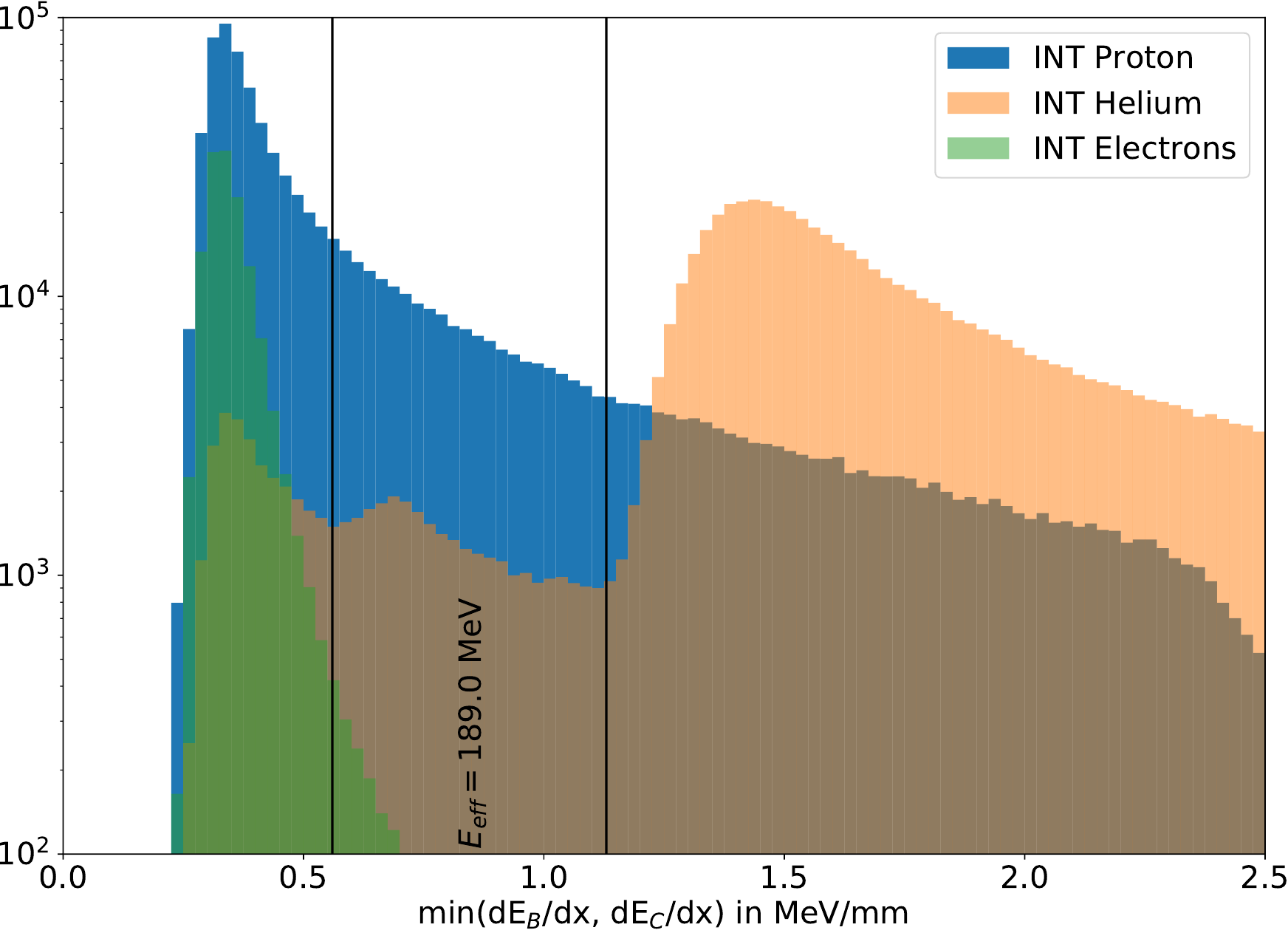}
    \caption{Contributions of electrons (green), protons (blue), and helium (orange) as functions of the minimum energy loss per path length in SSDs B and C, allowing for the minimisation of helium and electron contamination in the $>$100~MeV integral proton channel (see text for details).}
    \label{fig:dEdx-contributions}
\end{figure}

The helium-to-proton and electron-to-proton scaling was chosen 0.1 and arbitrarily, respectively. The latter takes into account the large variability of their relative intensities during SEP events \citep[see e.g.][]{Dresing-etal-2024}. Figure~\ref{fig:dEdx-contributions} shows the resulting contributions from electrons (green), protons (blue), and helium (orange).

For $\tfrac{dE}{dx} \gtrsim 1.13$~MeV/mm (second vertical line in Fig.~\ref{fig:dEdx-contributions}), the distributions of sub-100~MeV protons and several-hundred-MeV/amu helium become ambiguous until $\tfrac{dE}{dx} \gtrsim 2.5$~MeV/mm, where helium dominates.
At low $\tfrac{dE}{dx}$, electrons and protons overlap: below 0.56~MeV/mm (first vertical line), measurements can be dominated by $>$6~MeV electrons during SEP periods, while above 1.13~MeV/mm they can be influenced by relativistic helium during quiet times \citep[][and references therein]{Kuehl-etal-2015, Kuehl-etal-2016}.
The two vertical lines in the figure delimit the $\tfrac{dE}{dx}$ range selected for the $>$100~MeV proton channel. Note, the contribution induced by helium does not vanish as expected. The reason for these entries is hadronic interactions in the spacecraft producing secondary protons and deuterium, respectively.
Accordingly, the flux of $>$100~MeV proton proxy is estimated using events with $\tfrac{dE}{dx}$ between 0.56 and 1.13~MeV/mm.

\begin{figure}
    \centering
    \includegraphics[width=0.54\columnwidth]{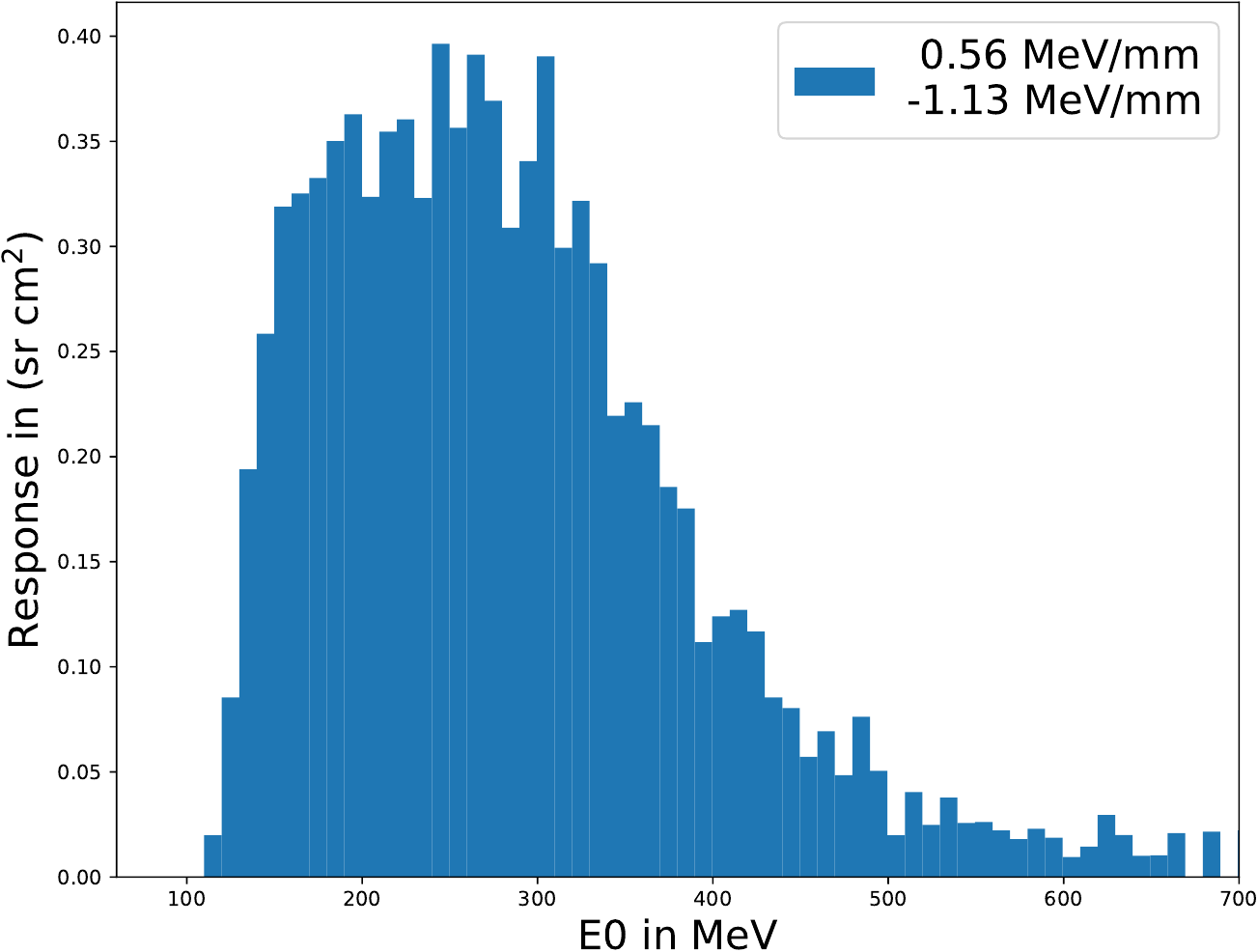}
    \includegraphics[width=0.44\columnwidth]{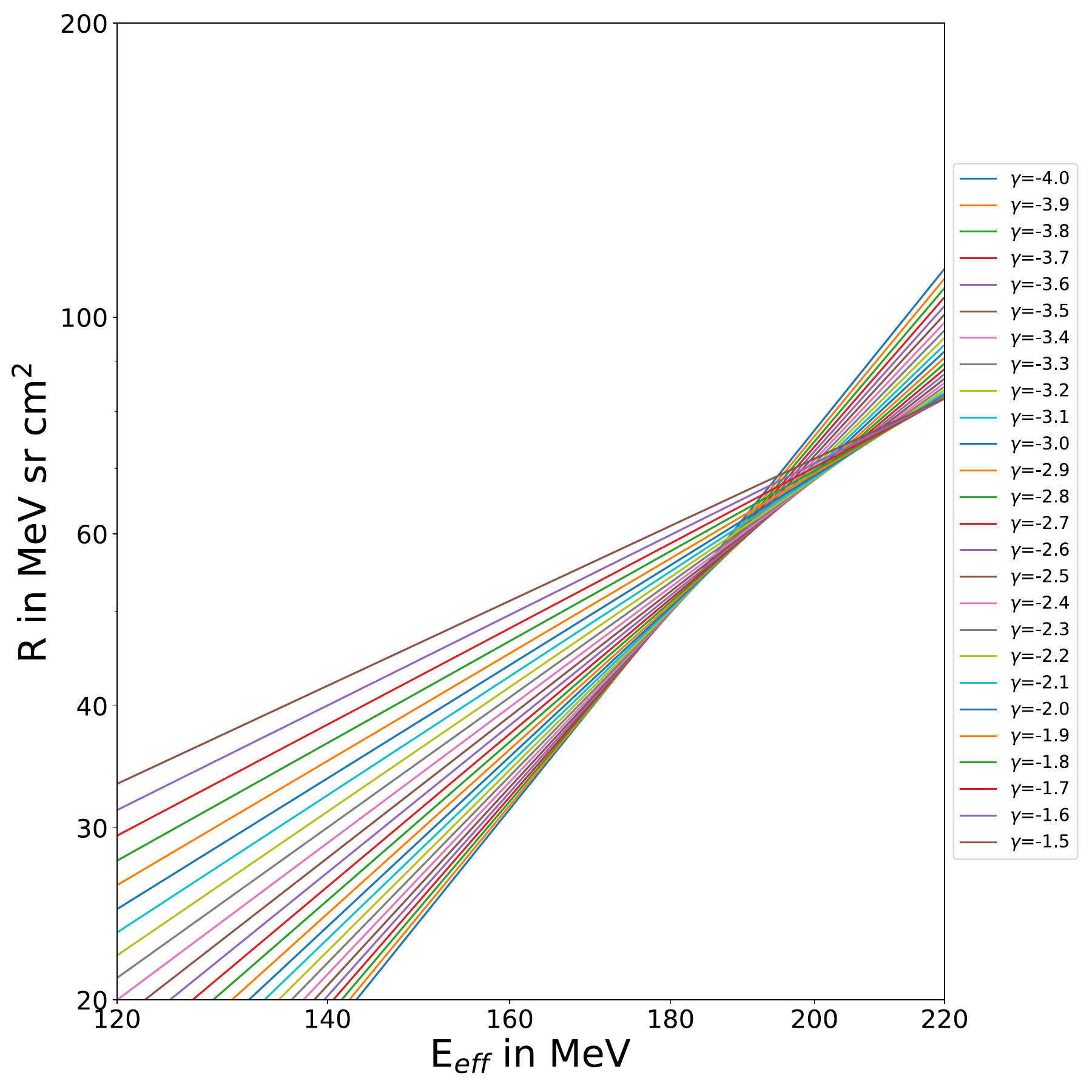}
    \caption{Left: Computed response function of the EPHIN $>$100~MeV proton channel. The effective energy, $E_{\rm eff}$, was determined using the bow-tie method \citep[][Sect. 5.3.3]{Oleynik_2021} with power laws of spectral index between $\gamma=2$ and $\gamma=5$. Only the central response is considered to minimise instrumental effects occurring at high flux levels.}
    \label{fig:weigthed-energy-gfac}
\end{figure}

Utilising these energy loss limits the energy response of the $>$100~MeV proton proxy was determined from the GEANT4  computations. The result is shown in the left panel of Fig.~\ref{fig:weigthed-energy-gfac}. indicating that the proxy is primarily sensitive to protons with energies between about 120 and 450~MeV. The flux, $\Phi$, can then be computed from the measured count rate, $C_{sel}$, in the penetrating channel fulfilling the boundary condition \[0.56 \leq \left(\tfrac{dE}{dx}/\tfrac{MeV}{mm}\right)_{B,C} \leq 1.13\] by 
\begin{equation} \label{eq:eq-sulivan}
    \Phi (E_{eff}) = \tfrac{C_{sel}}{R_{eff}} 
\end{equation}
A bow-tie analysis, as described, for example, in \citet{Oleynik_2021}, was applied to determine the effective energy $E_{\rm eff}$ and the corresponding integrated response function $R_{eff}$. The method used here relies on the assumption that the incoming proton spectra are power-law spectra $J(E,t)=J_0\cdot E^{\gamma}$ with indices between $-2$ and $-5$. Utilising the energy response function shown in the left panel of Fig.~\ref{fig:weigthed-energy-gfac} allows us to obtain $E_{\rm eff} = 189$~MeV and $R_{eff} = 60.34$~cm$^2$.sr.MeV. Note, $E_{\rm eff}$ obtained by EPHIN is somehow higher than the efficient energy from ERNE ranging between 128 and 172 MeV. 
 
\subsection{Fluence computation}
\begin{figure}
    \centering
    \includegraphics[width=0.95\linewidth]{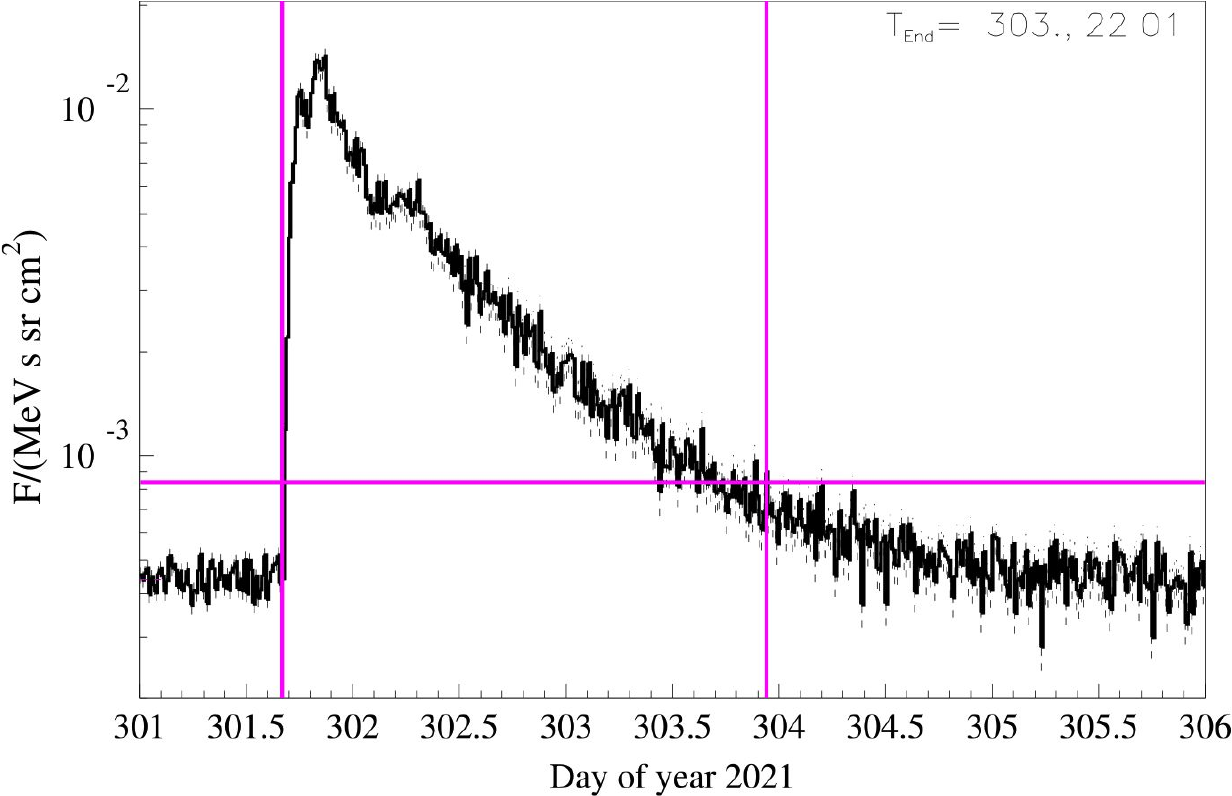}
    \caption{Example of fluence calculation from SOHO/EPHIN. The vertical magenta lines indicate the start and end times of the integration, while the horizontal line marks the background level used.}
    \label{fig:ephin_gle73}
\end{figure}

Figure~\ref{fig:ephin_gle73} illustrates the computation of the fluence for GLE73, showing the $>$100~MeV proton flux derived from EPHIN measurements.
Vertical magenta lines mark the start and end times of the integration, while the horizontal magenta line indicates the selected background level.

\begin{figure}
    \centering
    \includegraphics[width=0.99\linewidth]{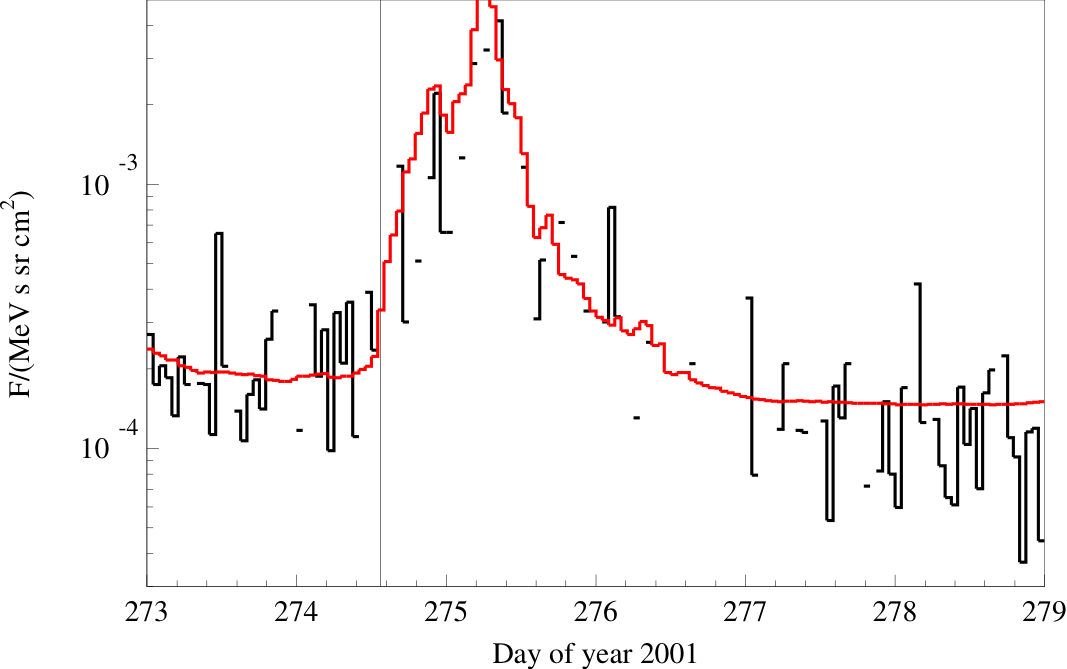}
    \includegraphics[width=0.99\linewidth]{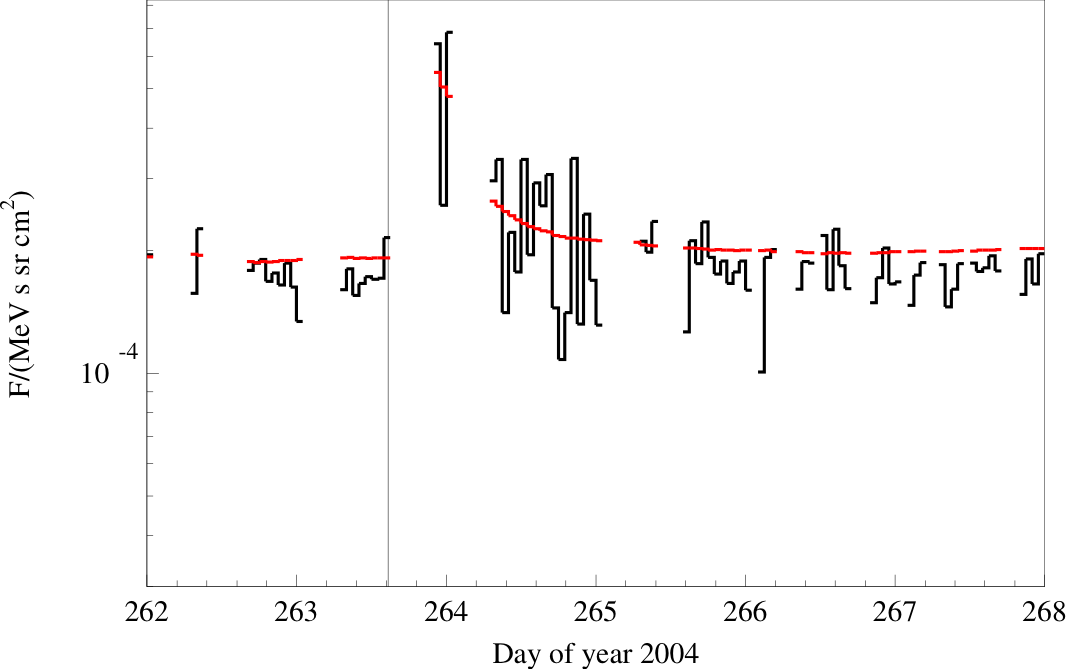}
    \caption{Top: Example of ‘low energy-loss statistics' (LELS) during the 1 October 2001 event.
    Bottom: Example of a ‘keyhole period' (KHP) on 20 September 2004.
    The black and red curves show the hourly averaged flux of the $>$100~MeV proton channel and the single-detector count rate of SSD F, respectively.
    The vertical lines mark the onset times determined from ERNE data.}
    \label{fig:EPHIN-low-statistics}
\end{figure}

\subsection{Limitations}
Figure~\ref{fig:EPHIN-low-statistics} shows two examples of limitations affecting the $>$100~MeV proton channel derived from SOHO/EPHIN.
The top panel illustrates the 1 October 2001 event, where data gaps result from limited telemetry and LELS issues, while the bottom panel shows the 20 September 2024 event, which occurred during a SOHO keyhole period.
In both cases, the vertical line marks the ERNE-determined onset time.
The black and red curves represent the hourly averaged flux of the $>$100~MeV proton channel and the single-detector count rate of SSD F, respectively — the latter serving as a proxy for $>$50~MeV protons, $>$50~MeV/amu helium, and $>$6~MeV electrons.

\section{Structure of the catalogue}
\label{annex: structure_catalogue}

\begin{itemize}

\item Columns 1 and 2: identification of the event
    \begin{itemize}
    \item Column 1: Catalogue entry number
    \item Column 2: Event identifier
    \end{itemize}

\item Columns 3 and 4: SOHO/ERNE data information
    \begin{itemize}
    \item Column 3: Time of onset (in UT)
    \item Column 4: Background mean of the channel [counts]
    \end{itemize}

\item Columns 5 to 11: SOHO/EPHIN data information
    \begin{itemize}
    \item Column 5: Background of the of $>$100 MeV proton proxy channel [1/(cm$^2$ sr s MeV)]
    \item Column 6: Peak flux of the event [1/(cm$^2$ sr s MeV)]
    \item Column 7: Relative uncertainty on peak flux [\%]
    \item Column 8: Fluence of the event [1/(cm$^2$ sr MeV)]
    \item Column 9: Relative uncertainty on fluence [\%]
    \item Column 10: End time (for the integration) (in DOY)
    \item Column 11: Comments
    \end{itemize}

\item Columns 12 to 19: soft X-rays flare information
    \begin{itemize}
    \item Column 12: Time of onset (in UT)
    \item Column 13: Time of peak flux (in UT)
    \item Column 14: Time of end (in UT)
    \item Column 15: Soft X-ray class (1--8 \AA~soft X-ray peak intensity in terms of C, M and X nomenclature where these represent intensities of 10$^{-6}$, 10$^{-5}$, and 10$^{-4}$ W/m$^2$, respectively)
    \item Column 16: Fluence [J/m$^2$]
    \item Column 17: X-ray location on the solar surface (in Stonyhurst heliographic co-ordinates, North (N) or South (S), and East (E) or West (W)) [degrees]
    \item Column 18: Spacecraft observer
    \item Column 19: Comments
    \end{itemize}

\item Columns 20 to 28: hard X-ray flare information
    \begin{itemize}
    \item Column 20: Time of onset (in UT)
    \item Column 21: Time of peak flux (in UT)
    \item Column 22: Time of end (in UT)
    \item Column 23: Peak intensity [counts/s]
    \item Column 24: Total counts [counts]
    \item Column 25: Energy range [keV]
    \item Column 26: Hard X-ray flare location on the solar surface (longitude, latitude in Carrington co-ordinates) [degrees]
    \item Column 27: Spacecraft/instrument observer
    \item Column 28: Comments
    \end{itemize}

\item Columns 29 to 35: radio type III information
    \begin{itemize}
    \item Column 29 to 31: Ground-based observations
    \begin{itemize}
        \item Column 29: Ground-based station observer
        \item Column 30: Onset time (in UT)
        \item Column 31: Frequency range [MHz]
    \end{itemize}
    \item Column 32 to 34: Space-borne observations
    \begin{itemize}
        \item Column 32: Spacecraft/instrument observer
        \item Column 33: Onset time (in UT)
        \item Column 34: Frequency range [kHz]
    \end{itemize}
    \item Column 35: Comments
    \end{itemize}

\item Columns 36 to 42: radio type II information
    \begin{itemize}
    \item Column 36 to 38: Ground-based observations
    \begin{itemize}
        \item Column 36: Ground-based station observer
        \item Column 37: Onset time (in UT)
        \item Column 38: Frequency range [MHz]
    \end{itemize}
    \item Column 39 to 41: Space-borne observations
    \begin{itemize}
        \item Column 39: Spacecraft/instrument observer
        \item Column 40: Onset time (in UT)
        \item Column 41: Frequency range [kHz]
    \end{itemize}
    \item Column 42: Comments
    \end{itemize}

\item Columns 43 to 49: radio type IV information
    \begin{itemize}
    \item Column 43 to 45: Ground-based observations
    \begin{itemize}
        \item Column 43: Ground-based station observer
        \item Column 44: Onset time (in UT)
        \item Column 45: Frequency range [MHz]
    \end{itemize}
    \item Column 46 to 49: Space-borne observations
    \begin{itemize}
        \item Column 46: Spacecraft/instrument observer
        \item Column 47: Onset time (in UT)
        \item Column 48: Duration [min]
        \item Column 49: End frequency [MHz]
    \end{itemize}
    \end{itemize}

\item Columns 50 to 56: gamma-rays information
    \begin{itemize}
    \item Column 50: Time of onset (in UT)
    \item Column 51: Time of peak flux (in UT)
    \item Column 52: Time of end (in UT)
    \item Column 53: Peak intensity [counts/s]
    \item Column 54: Total counts [counts]
    \item Column 55: Spacecraft/instrument observer
    \item Column 56: Comments
    \end{itemize}

\item Columns 57 to 61: Coronal mass ejection information
    \begin{itemize}
    \item Column 57: Time of first appearance (in UT)
    \item Column 58: Linear speed [km/s]
    \item Column 59: Space speed [km/s]
    \item Column 60: Angular width [deg]
    \item Column 61: Source of information
    \end{itemize}

\item Column 62 to 65: GLE information
    \begin{itemize}
    \item Column 62: GLE name
    \item Column 63: Integral intensity [\%.h]
    \item Column 64: Omni-directional integral fluence [cm$^{-2}$]
    \item Column 65: Name of observer
    \end{itemize}

\item Column 66: General notes

\end{itemize}

\section{Solar associations timings distribution}
\label{annex: timings_distrubtion}

\begin{table}[h!]
\caption{Statistics on timing delays between SEP release times and their associated solar phenomena.}
    \centering
    \renewcommand{\arraystretch}{1.12}
    \resizebox{\columnwidth}{!}{
    \begin{tabular}{@{} l c cc cc @{}}
        \toprule
        \multicolumn{1}{c}{$\Delta$} & $N$ & \multicolumn{2}{c}{Short} & \multicolumn{2}{c}{Long} \\
        \cmidrule(lr){3-4}\cmidrule(lr){5-6}
         &  & $>0$ & $<0$ & $>0$ & $<0$ \\
        \midrule
        $t_{\rm{SEP}}-t_{\rm{CME}}$ & 151 & 103 (68\%) & 48 (32\%) & 69 (46\%) & 82 (54\%) \\
        $t_{\rm{SEP}}-t_{\rm{SXR, onset}}$ & 121 & 119 (99\%) & 2 (1\%) & 79 (65\%) & 42 (35\%) \\
        $t_{\rm{SEP}}-t_{\rm{SXR, peak}}$ & 121 & 94 (78\%) & 27 (22\%) & 56 (46\%) & 65 (54\%) \\
        $t_{\rm{SEP}}-t_{\rm{HXR, onset}}$ & 87 & 76 (87\%) & 11 (13\%) & 53 (61\%) & 34 (39\%) \\
        $t_{\rm{SEP}}-t_{\rm{HXR, peak}}$ & 87 & 72 (83\%) & 15 (17\%) & 40 (46\%) & 47 (54\%) \\
        $t_{\rm{SEP}}-t_{\rm{tII}}$ & 126 & 91 (72\%) & 35 (28\%) & 60 (48\%) & 66 (52\%) \\
        $t_{\rm{SEP}}-t_{\rm{tIII}}$ & 155 & 144 (93\%) & 11 (7\%) & 99 (64\%) & 56 (36\%) \\
        $t_{\rm{SEP}}-t_{\rm{tIV}}$ & 40 & 25 (63\%) & 15 (37\%) & 17 (43\%) & 23 (57\%) \\
        $t_{\rm{SEP}}-t_{\gamma}$ & 57 & 51 (89\%) & 6 (11\%) & 37 (65\%) & 20 (35\%) \\
        \bottomrule
    \end{tabular}
    }
    \tablefoot{The first column indicates the delay considered, the second column gives the number of events in the statistics. In the third and fourth columns, we consider the short and long travel times for the particles (see Sect. \ref{sec: results} for details), and display the number of events with positive or negative delays. Percentages are relative to the total number of SEP events with an identified solar association.}
    \label{tab:statistics_solar_assoc}
\end{table}

The most significant timing differences ($>$360 minutes) correspond to the following SEP events (from largest to smallest):
\begin{itemize}
    \item Event 110 (6 March 2013, $\Delta(t_{SEP}-t_{CME}) = 1237$ minutes), which presents a very slow rising phase
    \item Event 114 (11 October 2013, $\Delta(t_{SEP}-t_{CME}) = 831$ minutes), which also presents a very slow rising phase
    \item Event 61 (8 November 2003, $\Delta(t_{SEP}-t_{CME}) = 831$ minutes), for which only one strong far side CME occurring on the 7 November 2003 seems to be relevant. A more realistic onset could be hidden in the decay phase of the previous $>$100~MeV proton event (number 60, 4 November 2003)
    \item Event 89 (5 June 2011, $\Delta(t_{SEP}-t_{CME}) = 790$ minutes), associated with a strong CME on the 4 June 2011 but which could also be related to a smaller one occurring at 08:00 UTC on 5 June 2011,  enriched in high-energy SEPs due to the previous event
    \item Event 76 (14 July 2005, $\Delta(t_{SEP}-t_{CME}) = 626$ minutes), associated to a strong CME but alternatively related to two other smaller CMEs which occurred, respectively, six and seven hours after the first one
    \item Event 124 (1 September 2014, $\Delta(t_{SEP}-t_{CME}) = 453$ minutes)
    \item Event 167 (13 May 2024, $\Delta(t_{SEP}-t_{CME}) = 452$ minutes)
\end{itemize}

\onecolumn
\section{Radio instruments}
\label{annex: radio_instruments}

\begin{table*}[ht!]
\caption{Space-borne and ground-based instruments used in the search for radio burst associations with SEP events in the $>$100 MeV proton catalogue, along with their corresponding frequency ranges and operational periods.}
  \begin{center}
    \begin{tabular}{l|c|c|c|l} 
      \multicolumn{5}{c}{\textbf{Ground-based instruments}}\\
    \hline
     \textbf{Station} & \textbf{Acronym} & \textbf{Frequency range (MHz)} & \textbf{Timespan} & \textbf{Location}\\
      \hline
      Learmonth & LEAR & 25-180 & 1975-present & Learmonth, Australia \\
      San Vito & SVTO & 25-180 & 1994-present & San Vito, Italy \\
      Sagamore Hill & SGMR & 25-180 & 1975-present & Massachusetts, USA \\
      Potsdam & POTS & 40-800 & 1993-2006 & Potsdam, Germany \\
      Holloman & HOLL & 25-180 & 1996-present & New Mexico, USA \\
      Palehua & PALE & 25-180 & 1980-present & Hawaii, US \\
      Culgoora & CULG & 18-1800 & 1975-2019 & New South Wales, Australia \\
      HiRAS & HIRA & 25-2500 & 1992-2016 & Hiraiso, Japan \\
      Izmiran & IZMI & 20-260 & 1985-2016 & Moscow, Russia \\
      Nancay Decameter Array & NDA & 10-120 & 1980-present & Nancay, France \\
      ORFEES & ORFEES & 144-1004 & 1975-present & Nancay, France \\
      ARTEMIS-IV & ARTEMIS & 20-650 & 1998-2013 & Thermopylae, Greece \\
      Bleien & BLEN & 100–1000 & 1979-2014 & Aargau, Switzerland \\
      \\
    
       \multicolumn{5}{c}{\textbf{Space-borne instruments}}\\
      \hline
      \textbf{Spacecraft} & \textbf{Instrument} & \textbf{Frequency range (MHz)} & \textbf{Timespan} & \textbf{Orbit}\\
      \hline
      WIND & WAVES & 0.02-13.8 & 1994-present & L1 \\
      STEREO A & WAVES & 0.01-16.1 & 2006-present & Heliocentric $\backsim$1 AU \\
      STEREO B & WAVES & 0.01-16.1 & 2006-2014 & Heliocentric $\backsim$1 AU \\
      \hline
      \label{tab:radio_instrum}
    \end{tabular}
  \end{center}
\end{table*}

\FloatBarrier
\twocolumn


\end{document}